\documentclass[twoside,twocolumn,aps,10pt,prb,floatfix,showpacs,citeautoscript,superscriptaddress,longbibliography]{revtex4-2}

\usepackage{amsmath}
\usepackage{amssymb}
\usepackage{graphicx}
\usepackage{tabularx}
\usepackage{dcolumn}
\usepackage{xcolor}
\usepackage{booktabs}
\usepackage{comment}
\usepackage{enumitem}
\usepackage{multirow}
\usepackage[acronym]{glossaries}
\usepackage[version=4]{mhchem}

\usepackage{siunitx}
\DeclareSIUnit\angstrom{\text {Å}}
\definecolor{cadmiumgreen}{rgb}{0.0, 0.42, 0.24}
\newcommand{\BZO}{BaZrO$_\mathrm{3}$~}

\usepackage{braket}

\usepackage[colorlinks=true]{hyperref}
\hypersetup{
    pdftoolbar=true,        
    pdfmenubar=true,        
    pdffitwindow=false,     
    pdfstartview={FitH},    
    pdftitle={My title},    
    pdfauthor={Author},     
    pdfsubject={Subject},   
    pdfcreator={Creator},   
    pdfproducer={Producer}, 
    pdfkeywords={keyword1, key2, key3}, 
    pdfnewwindow=true,      
    colorlinks=true,        
    linkcolor=black,      
    citecolor=blue,        
    filecolor=blue,      
    urlcolor=blue        
}

\graphicspath{{/article-v2/figures/}} 

\setacronymstyle{long-short}

\newacronym{aimd}{AIMD}{ab-initio molecular dynamics}
\newacronym{ardr}{ARDR}{automatic relevance detection regression}
\newacronym{bcc}{BCC}{body-centered cubic}
\newacronym{cv}{CV}{cross-validation}
\newacronym{ce}{CE}{cluster expansion}
\newacronym{dft}{DFT}{density functional theory}
\newacronym{dof}{DOF}{degrees of freedom}

\newacronym{dos}{DOS}{density of states}
\newacronym{edos}{EDOS}{electronic density of states}
\newacronym{vdos}{VDOS}{vibrational density of states}

\newacronym{eam}{EAM}{embedded atom method}
\newacronym{eci}{ECI}{effective cluster interaction}
\newacronym{emt}{EMT}{effective medium theory}
\newacronym{ehm}{EHM}{effective harmonic model}
\newacronym{ha}{HA}{harmonic approximation}
\newacronym{qha}{QHA}{quasi-harmonic approximation}
\newacronym{fc}{FC}{force constant}
\newacronym{fcc}{FCC}{face-centered cubic}
\newacronym{fcp}{FCP}{force constant potential}
\newacronym{gc}{GC}{grand canonical}
\newacronym{lasso}{LASSO}{least absolute shrinkage and selection operator}
\newacronym{loocv}{LOOCV}{leave-one-out cross-validation}
\newacronym{mae}{MAE}{mean absolute error}
\newacronym{mc}{MC}{Monte Carlo}
\newacronym{md}{MD}{molecular dynamics}
\newacronym{msd}{MSD}{mean squared displacement}
\newacronym{msrd}{MSRD}{mean squared relative displacement}
\newacronym{ols}{OLS}{ordinary least squares}
\newacronym{omp}{OMP}{orthogonal matching pursuit}
\newacronym{pes}{PES}{potential energy surface}
\newacronym{rfe}{RFE}{recursive feature elimination}
\newacronym{rmse}{RMSE}{root-mean-square error}
\newacronym{scp}{SCP}{self-consistent phonon}
\newacronym{hp}{HP}{harmonic phonon}
\newacronym{sgc}{SGC}{semi-grand canonical}
\newacronym{sed}{SED}{spectral energy density}

\newacronym{svd}{SVD}{singular value decomposition}
\newacronym{tmd}{TMD}{transition metal di\-chal\-co\-ge\-ni\-de}
\newacronym{vacf}{VACF}{velocity auto-correlation function}
\newacronym{fep}{FEP}{free energy perturbation}
\newacronym{tdep}{TDEP}{temperature dependent effective potential}

\newcommand{\hiphive}{\textsc{hiphive}}

\newcommand{\phonopy}{\textsc{phonopy}}

\newcommand{\dynasor}{\textsc{dynasor}}
\newcommand{\vasp}{\textsc{vasp}}

\renewcommand{\epsilon}[0]{\varepsilon}

\newcommand{\eq}[1]{Eq.~\eqref{#1}}

\makeatletter
\g@addto@macro\bfseries{\boldmath}
\makeatother

\begin{document}



\title{Anharmonicity of the antiferrodistortive soft mode in barium zirconate \texorpdfstring{\ce{BaZrO3}}{BaZrO3}}

\author{Petter Rosander}
\author{Erik Fransson}
\affiliation{Department of Physics, Chalmers University of Technology, SE-412 96  G\"oteborg, Sweden}

\author{Cosme Milesi-Brault}
\affiliation{Department of physics and materials science, University of Luxembourg, 41 Rue du Brill, L-4422 Belvaux, Luxembourg}
\affiliation{Materials Research and Technology Department, Luxembourg Institute of Science and Technology, 41 rue du Brill, L-4422 Belvaux, Luxembourg}
\affiliation{Institute of Physics of the Czech Academy of Sciences, Na Slovance 1999/2, 182 21 Prague, Czech Republic}
\author{Constance Toulouse}
\affiliation{Department of physics and materials science, University of Luxembourg, 41 Rue du Brill, L-4422 Belvaux, Luxembourg}



\author{Fr\'ed\'eric Bourdarot}
\affiliation{Service de Mod\'elisation et d’Exploration des Mate\'eriaux, Universit\'e Grenoble Alpes et Commissariat \`a l’\'Energie Atomique, INAC, 38054 Grenoble, France}
\author{Andrea Piovano}
\affiliation{Institut Laue-Langevin (ILL), 71 avenue des Martyrs, 38042 Grenoble, France}
\author{Alexei Bossak}
\affiliation{European Synchrotron Radiation Facility, BP 220, 38043 Grenoble, France}

\author{Mael Guennou}
\email{mael.guennou@uni.lu}
\affiliation{Department of physics and materials science, University of Luxembourg, 41 Rue du Brill, L-4422 Belvaux, Luxembourg}

\author{G\"oran Wahnstr\"om}
\email{goran.wahnstrom@chalmers.se}
\affiliation{Department of Physics, Chalmers University of Technology, SE-412 96  G\"oteborg, Sweden}


\date{\today}

\begin{abstract}
Barium zirconate (BaZrO$_3$) is one of the very few perovskites that is claimed to retain an average cubic structure down to \SI{0}{\K}, while being energetically very close to an antiferrodistortive phase obtained by condensation of a soft phonon mode at the R point of the Brillouin zone boundary.
In this work, we report a combined experimental and theoretical study of the temperature dependence of this soft phonon mode.
Inelastic neutron and x-ray scattering measurements on single crystals show that it softens substantially from \SI{9.4}{\meV} at room temperature to \SI{5.6}{\meV} at \SI{2}{\K}. In contrast, the acoustic mode at the same R point is nearly temperature independent.
The effect of the anharmonicity on the lattice dynamics is investigated non-perturbatively using direct dynamic simulations as well as a first-principles based self-consistent phonon theory, including quantum fluctuations of the atomic motion.
By adding cubic and quartic anharmonic force constants, quantitative agreement with the neutron data for the temperature dependence of the antiferrodistortive mode is obtained.
The quantum fluctuations of the atomic motion are found to be important to obtain the proper temperature dependence at low temperatures.
The mean squared displacements of the different atoms are determined as function of temperature and are shown to be consistent with available experimental data.
Adding anharmonicity to the computed fluctuations of the Ba-O distances also improves the comparison with available EXAFS data at \SI{300}{\K}.

\end{abstract}

\maketitle

\clearpage

\section{Introduction}
\label{sec:introduction}

Perovskite oxides constitute a prominent class of materials with a wide range of different properties such as ferroelectricity, colossal magnetoresistance, electronic and/or ionic conductivity,  piezoelectricity, superconductivity, metal-insulator transition, luminescence, and much more \cite{bhallaPerovskiteStructureReview2000}. The ideal perovskite structure is cubic, with the general chemical formula ABO$_3$, where the A and B sites can accommodate a wide variety of elements from the periodic table. Many perovskites are cubic at high temperatures but upon cooling most undergo one or several structural phase transitions, which depend sensitively on the choice of A and B \cite{glazerBriefHistoryTilts2011,howardStructuresPhaseTransitions2005}. This variety of compositions and structural phases yields the wide range of properties observed in these materials. The two most important structural distortions are the off-centering of the B-site cation, giving rise to ferroelectricity,
and the tilts or rotations of the BO$_6$ octahedra
\cite{saiFirstprinciplesStudyFerroelectric2000}.
The former is detectable as an instability at the $\Gamma$ point in the Brillouin zone, while the latter are related to instabilities at the R and/or M points of the Brillouin zone boundary. To understand, predict and control these structural distortions is a key issue in the research on perovskite oxides.

Barium zirconate BaZrO$_3$ (BZO) is a simple perovskite that has attracted considerable interest due to its excellent thermal stability, high melting temperature, low thermal expansion coefficient, low dielectric loss, and
low thermal conductivity \cite{knightLowtemperatureThermophysicalCrystallographic2020}.
These properties make BZO highly attractive for a range of technological applications such as wireless communication \cite{sebastian2010dielectric}, thermal barrier coatings for gas turbines \cite{vassenZirconatesNewMaterials2000}, protonic fuel cell applications \cite{KreuerProtonConductingOxides2003} and ceramic reactors \cite{clarkSinglestepHydrogenProduction2022}, as well as substrates in thin film deposition \cite{dobalMicroRamanScatteringDielectric2001}.

Barium zirconate (BZO) is also one of the very few perovskites that is claimed to remain cubic down to 0 K, but its tendency towards an antiferrodistortive instability has fueled a lot of studies and debates about its ground state, its average structure, and the possibility for local disorder. 
Several first-principles calculations, based on the density functional theory (DFT), predict the cubic structure to be unstable at \SI{0}{\K}, due to a soft phonon mode at the R point of the Brillouin zone corresponding to antiphase tilts of sequential oxygen octahedra, {\it i.e.} an antiferrodistortive mode \cite{zhongCompetingStructuralInstabilities1995,bennettEffectSymmetryLowering2006,akbarzadeh_combined_2005,bilicGroundStateStructure2009,chenEnergeticsOxygenoctahedraRotations2018,Amoroso2018}.
However, the DFT results are sensitive to the exchange-correlation functional \cite{perrichonUnravelingGroundStateStructure2020} and by using hybrid functionals a stable cubic structure is obtained at \SI{0}{\K}
\cite{evarestovHybridDensityFunctional2011,perrichonUnravelingGroundStateStructure2020}.
It has been suggested that, in cases where the instability, as observed in the DFT calculations, is weak, quantum fluctuations may suppress the phase transition in BZO and the material therefore stays cubic all the way down to \SI{0}{\K}
\cite{zhongCompetingStructuralInstabilities1995,akbarzadeh_combined_2005,chenEnergeticsOxygenoctahedraRotations2018}.
Experimentally, both X-ray and neutron diffraction data on a powder sample, suggest a cubic structure down to T = 2 K \cite{akbarzadeh_combined_2005}, which is further supported by Raman spectroscopy on a single crystal \cite{toulouse2019}. 
On the other hand, a recent study combining electron diffraction with total neutron scattering measured on ceramic samples claimed that the material underwent a structural change below 80 K associated with the onset of correlated tilts of ZrO$_6$ octahedra \cite{levinNanoscalecorrelatedOctahedralRotations2021}.

Clearly, direct measurements and theoretical modeling of the tilt mode frequency and its temperature dependence are essential for a good understanding of BZO.
Experimentally, direct lattice dynamical studies have been limited by the scarcity of suitable large and high-quality single crystals, as their synthesis is made difficult by the very high melting point of BZO.
As a result, the only experimental study of the temperature dependence of the R-tilt mode of BZO has been attempted on a polycrystalline sample \cite{perrichonUnravelingGroundStateStructure2020}; it concluded that the tilt mode lies at 5.9 meV at low temperatures and does not exhibit any significant temperature dependence between 5 and \SI{500}{\K}. Recently high-quality BZO single crystals have been synthesized using the floating zone technique \cite{xin_single_2019}. Here we take advantage of those single crystals to investigate directly the low-frequency lattice dynamics. 

Quantitative theoretical modelling of the temperature dependence of the low frequency vibrational motion of BZO is challenging.
In recent years several methods have been developed to treat vibrational anharmonicities for solids, such as the self-consistent ab-initio lattice dynamics (SCAILD) technique \cite{souvatzisEntropyDrivenStabilization2008}, the temperature dependent effective potential (TDEP) method \cite{hellmanLatticeDynamicsAnharmonic2011}, and the stochastic self-consistent harmonic approximation (SSCHA) \cite{monacelliStochasticSelfconsistentHarmonic2021}.
Methods based on the self-consistent phonon approach \cite{liaoNanoscaleEnergyTransport2020} have also been developed by Tadano et al. \cite{tadanoSelfconsistentPhononCalculations2015}, which implements the Green's function technique to derive the anharmonic contributions.
Recently, this latter technique was used to study anharmonicity-induced phonon hardening and thermal transport in BZO in the temperature range \SI{300}{K} - \SI{2000}{K} \cite{Zheng2022}.

Here, a force constant potential (FCP), expanded up to fourth order, is derived
based on training structures from \gls{dft} calculations and by making use of regression techniques \cite{EriFraErh19}.
The FCP takes anharmonicity into account and it is used in molecular dynamics (MD) simulations \cite{larsenAtomicSimulationEnvironment2017,fanEfficientMolecularDynamics2017} to capture the full dynamical structure factor and its temperature dependence \cite{FraSlaErhWah2021}.
A self-consistent phonon (SCP) approach is then implemented \cite{EriFraErh19} to, in addition to anharmonicity, also include the quantum fluctuations of the atomic motion, known to be significant at low temperatures \cite{akbarzadeh_combined_2005}.

The paper is organized as follows. 
In Sec.~\ref{sec:methods} the theoretical methods used to treat the anharmonic lattice vibrations are introduced, including the effect of the quantum fluctuations of the atomic motions at low temperatures.
Sec.~\ref{sec:exp_results} presents the experimental data obtained by inelastic neutron and x-ray scattering that reveal the strong softening of the tilt mode and the comparatively temperature independent behaviour of the acoustic mode. 
In Sec.~\ref{sec:theory_results}, the theoretical results are presented and compared with our experimental data. The strong temperature dependence of the soft tilt mode is found to be in quantitative agreement with our theoretical predictions and the obtained essentially temperature independent acoustic mode also agree with the theoretical predictions.
In Sec.~\ref{sec:theory_results_atomic_displacemnts} it is shown that our theoretical results for the atomic displacements agree favourably with available experimental data for the mean squared displacements as well as with the more local information obtained in extended X-ray absorption fine structure (EXAFS) spectroscopy.
Finally, Sec.~\ref{sec:discussion} discusses our present theoretical results and put them into a broader context and Sec.~\ref{sec:conclusions} summarizes our main conclusions.

\section{Theoretical methods}
\label{sec:methods}

\subsection{Density functional theory}
\label {subsec:DFT}

The \gls{dft} calculations are carried out using the Vienna ab-initio simulation package (\vasp\ \cite{KreFur1996-1,KreFur1996-2}).

The potential energy surface (PES) for the R-tilt mode depends quite sensitively on the exchange--correlation functional \cite{perrichonUnravelingGroundStateStructure2020}.
The local density approximation (LDA), two generalized gradient approximations (PBE \cite{PerBurErn1996} and PBEsol \cite{perdewRestoringDensityGradientExpansion2008} ) as well as a functional including non-local correlation effects (CX \cite{Berland_Hyldgaard_2014,dionVanWaalsDensity2004}) all give a double minimum PES for the R-tilt mode coordinate using the experimental value for the lattice constant.
Using the corresponding theoretical lattice constants, LDA gives deeper potential minima, while for PBE the double minimum feature disappears.
Two hybrid functionals (HSE \cite{Heyd_Scuseria_Ernzerhof_2003,Heyd_Scuseria_Ernzerhof_2006} and CX0p \cite{Jiao_Schröder_Hyldgaard_2018}) show in general better agreement for the various vibrational frequencies and both show an anharmonic PES for the R-tilt mode, quite similar to PBE at its theoretical lattice constant. The computational cost for the hybrid functionals is substantial compared with generalized gradient approximations and as a compromise between accuracy and computational cost we have therefore chosen to base this study on PBE using its theoretical lattice constant. For comparison, PBEsol has also been used, and the corresponding results are presented in \autoref{sec:discussion}.

The included projector-augmented wave (PAW) \cite{Blo1994,KreJou1999} potentials are used with energy cut-off \SI{500}{eV}.
The considered valence configurations for Ba, Zr and O are $5s^{2}5p^{6}6s^{2}$, $4s^{2}4p^{6}4d^{2}5s^{2}$ and $2s^{2}2p^{2}$ respectively. 
The Brillouin zone of the primitive cell is sampled with an 8x8x8 Monkhorst-Pack $k$-point mesh and scaled accordingly for larger cells.
The obtained lattice constants for PBE and PBEsol are $a$ = \SI{4.236}{\angstrom} and $a$ = \SI{4.192}{\angstrom}, respectively.

The non-analytic part of the dynamical matrix, which is due to the long-range dipole-dipole interaction \cite{PhysRevB.50.13035,gonzeDynamicalMatricesBorn1997}, is evaluated and applied in calculating the phonon dispersion relations using \phonopy{} \cite{TOGO20151}. 
The values for the dielectric tensor, $\epsilon$, and Born effective charges, $Z^*$, are computed to be $\epsilon=4.87$, $Z^*(\textrm{Ba})=2.72$, $Z^*(\textrm{Zr})=6.11$, 
and $Z^*(\textrm{O})_\parallel=-4.85$ and $Z^*(\textrm{O})_\perp=-1.99$ for the two different displacement directions of the oxygen ion.

\subsection{Force constant potential}

The force constant potential is based on an expansion of the potential energy $U$ in terms of atomic displacements $u_{i}^{\alpha}$ from their equilibrium positions, $\mathbf{R}^0_i$. We consider an expansion up to fourth order
\begin{equation}
   \begin{aligned}
    U = U_0 + \frac{1}{2!}&\Phi_{ij}^{\alpha\beta} u_{i}^{\alpha}u_{j}^{\beta} 
    + \frac{1}{3!}\Phi_{ijk}^{\alpha\beta\gamma} 
    u_{i}^{\alpha}u_{j}^{\beta}u_{k}^{\gamma} 
    \\
    &+ \frac{1}{4!}\Phi_{ijkl}^{\alpha\beta\gamma\delta} 
    u_{i}^{\alpha}u_{j}^{\beta}u_{k}^{\gamma}u_{l}^{\delta}
    \label{eq:fcp_expansion}
   \end{aligned}
\end{equation}
where the $\Phi$s are referred to as force constants (FCs). Latin indices run over atomic labels, Greek over Cartesian coordinates and Einstein summation applies. 

First, an initial phonon calculation is carried out for a 3x3x3 supercell (135 atoms) using \gls{dft}.
These phonons are used to construct a set of training structures with displacements generated via populating the normal modes with different temperatures. These structures are used to construct an initial fourth order \gls{fcp} using \hiphive{} \cite{EriFraErh19}.

Next, the size of the system is increased to a 4x4x4 supercell (320 atoms). The initial fourth order FC model is used to run \gls{md} simulations carried out at 10, 150, 300 and \SI{500}{K}.
Snapshots from these simulations together with some small amplitude structures along the R-tilt mode are used as training structures for the final model.
For the final model, we considered two and three body interactions up to, and including the fourth order.
We used a maximum interaction length of \SI{8}{\angstrom} for the harmonic force constants.
Whereas, for the anharmonic interactions, that is, the third and fourth order interaction, we set a maximum interaction length for the force constants to \SI{5}{\angstrom}.
The training was carried out using the recursive feature elimination method \cite{scikit-learn}.
We denote this final model as the {\bf force constant potential (FCP)} model.
The accuracy of the \gls{fcp} is evaluated using \gls{cv}, and an average  \gls{rmse} of \SI{0.034}{eV/\angstrom} is obtained over the validation set. 
For the CV-plot, see Fig.~S1 in Supplemental Material \cite{SI:this}.

\begin{figure}[ht]
\centering
\includegraphics{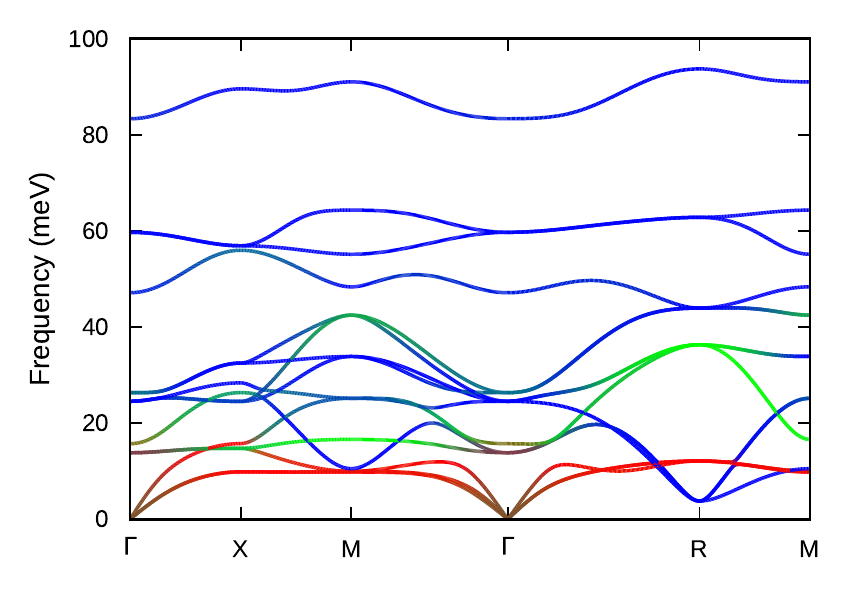}
\caption
{Phonon dispersion relations within the harmonic model (the HP model) for BZO using PBE. Color assigned as (Ba, Zr, O) = (red, green, blue).
}
\label{fig:harmonic_dispersion}
\end{figure}

\subsection{Harmonic phonon model}

By truncating the \gls{fcp} model at the second order term an harmonic approximation is obtained. We will denote this, the {\bf \gls{hp}} model. 
Within the harmonic approximation eigenfrequencies $\omega_{{\boldsymbol{q}},\nu}$ and eigenvectors $\boldsymbol{e}_{{\boldsymbol{q}},\nu}$ can be derived \cite{TOGO20151}.
Here, $\boldsymbol{q}$ is the wavevector and $\nu$ the band index. 
In \autoref{fig:harmonic_dispersion} we show the phonon dispersion for BZO using the HP model and where the correction due to the long-range dipole-dipole interaction \cite{PhysRevB.50.13035,gonzeDynamicalMatricesBorn1997} has been applied.
At the R-point the lowest branch at \SI{3.8}{\milli\electronvolt} corresponds to antiphase tilts of sequential ZrO$_6$ octahedra. This will here be denoted the {\bf R-tilt mode}.
The second lowest branch located at \SI{12.1}{\milli\electronvolt} at the R point, is the acoustic branch due to barium displacements, here denoted the {\bf R-acoustic mode}.

The phonon dispersions in \autoref{fig:harmonic_dispersion} are similar to what is obtained from the standard small displacement method (for a comparison, see Fig.~S2 in Supplemental Material \cite{SI:this}). However, the R-tilt mode frequency is sensitive to the theoretical treatment due to anharmonicity. Here, we have developed a harmonic model based on displacements from \gls{md} simulations carried out at 10, 150, 300 and \SI{500}{K}. The obtained R-tilt mode frequency is \SI{3.8}{\milli\electronvolt}, slightly larger than \SI{2.7}{\milli\electronvolt}, obtained using the standard small displacement method with the default displacement $\pm$ 0.01
(see Fig.~S2 in Supplemental Material \cite{SI:this}).

\begin{figure}[ht]
\centering
\includegraphics{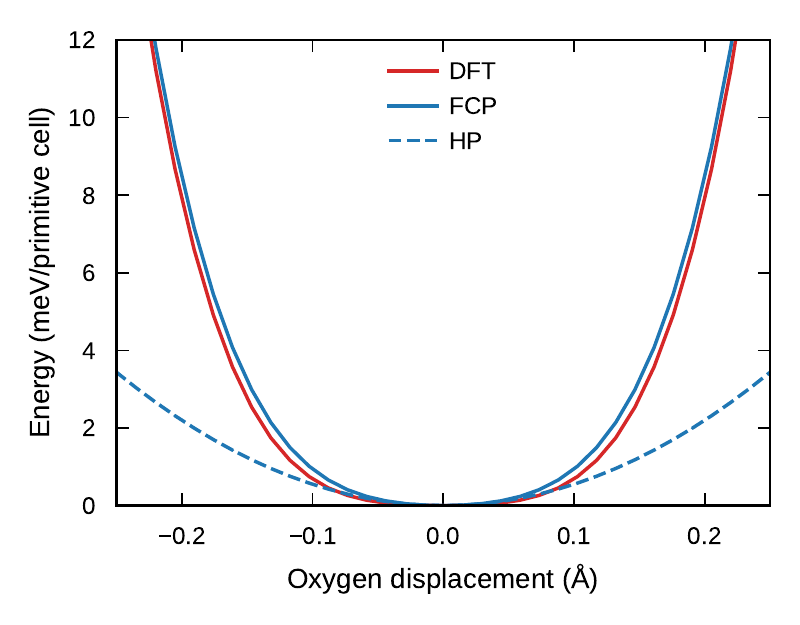}
\caption{
    The potential energy surface (PES) along the R-tilt mode based on PBE. The fourth order \gls{fcp} model agrees very well with the \gls{dft} data, while the second order model, the \gls{hp} model, deviates markedly. 
}
\label{fig:model_validation_Rmode_pes}
\end{figure}

The PES along the R-tilt mode is shown in \autoref{fig:model_validation_Rmode_pes}. The \gls{fcp} model agrees very well with the \gls{dft} data. In contrast, the \gls{hp} approximation deviates markedly from the \gls{dft} data, demonstrating that the R-tilt mode is strongly anharmonic.

Within the harmonic approximation the temperature dependent atomic displacement of atom $i$ in direction $\alpha$, $u_{i}^\alpha$, is given by a Gaussian distribution with mean value zero and variance
\begin{align}
   \langle (u_{i}^{\alpha})^2 \rangle = 
   \frac{1}{N_q} \sum_{{\boldsymbol{q}},\nu} 
   \frac{\hbar}{2 m_i \omega_{{\boldsymbol{q}},\nu}}
   \mid e_{{\boldsymbol{q}},\nu}^{i,\alpha} \mid^2
   \coth{(\hbar\omega_{{\boldsymbol{q}},\nu}/2k_{\text B}T)}
   \label{eq:u2_harm}
\end{align}
Here, $\langle \ldots \rangle$ denotes a thermal average, $m_i$ is the mass of atom $i$, and $N_q$ the number of $q$-values in the summation.

\subsection{Self-consistent phonon model}

In the {\bf self-consistent phonon (SCP)} model, a temperature dependent effective harmonic model is constructed. We start with the harmonic model. Displacements $u_i^{\alpha}$ are generated in a supercell using the Box-Muller method according to
\begin{align}
   u_{i}^{\alpha} = \sum_{\lambda} A_{\lambda}^{i} 
   e_{\lambda}^{i,\alpha}
   \sqrt{-2 \ln{U_{\lambda}}} \cos{(2 \pi Q_{\lambda})}
    \label{eq:u_displacements}
\end{align}
where the index $\lambda = (\boldsymbol{q},\nu)$ enumerates the points commensurate with the supercell, and ${U_{\lambda}}$ and ${Q_{\lambda}}$ are uniform random numbers on [0,1]. The temperature dependent amplitude $A_{\lambda}^{i}$ is given by the standard deviation of the displacements (cf. \eq{eq:u2_harm})
\begin{equation}
 A_{\lambda}^{i} = \sqrt{\hbar\coth{(\hbar\omega_{\lambda}/2k_{\text B}T)}/{(2 m_i \omega_{\lambda}})}
 \label{eq:amplitude_qm}
\end{equation}

Using \eq{eq:u_displacements} a set of displacements $\mathbf{u}$ is generated at a chosen temperature $T$ and the corresponding forces are calculated using the \gls{fcp}. We generate M such reference structures. The generated displacements and forces are encoded in the fit matrix ${\mathbf A}(\mathbf{u})$ and the force vector ${\mathbf f}(\mathbf{u})$, respectively. An effective harmonic model is obtained by solving the minimization problem 
\begin{equation}
    \min_{\mathbf{x}}
    \lVert {\mathbf{A}}(\mathbf{u}) {\mathbf x} - {\mathbf{f}}(\mathbf{u}) \rVert
    \label{eq:scph_minimization}
\end{equation}
where $\mathbf x$ are the free parameters of effective harmonic model. The procedure is iterated until convergence
\begin{equation}
    {\mathbf x}_{n+1} = (1 - \alpha)\ {\mathbf x}_{n}
    + \alpha \min_{\mathbf{x}}
    \lVert {\mathbf{A}}(\mathbf{u}_n) {\mathbf x} - {\mathbf{f}}(\mathbf{u}_n) \rVert
    \label{eq:scph}
\end{equation}
where $\mathbf{u}_n$ is a set of displacements generated with the $n$:th effective harmonic model ($\mathbf x_n$) and $\alpha$ is a suitable chosen parameter to obtain efficient numerical convergence \cite{EriFraErh19}. A self-consistent effective harmonic model at temperature $T$ is then obtained.

The above procedure is repeated at different temperatures and temperature dependent eigenfrequencies and eigenvectors are obtained.

\section{Experimental methods and results}
\label{sec:exp_results}

\subsection{Experimental details}

\BZO single crystals were grown by the optical floating zone technique as described in  Ref.~\cite{xin_single_2019}. A single crystal was extracted from the boule for inelastic neutron scattering (INS) measurements with irregular shape and some natural facets suitable for sample orientation, with a total mass of approximately \SI{200}{\milli\gram} and a size of a few millimetres. The sample for inelastic x-ray scattering (IXS) was cut from another smaller piece and thinned down to a lamella of approximately \SI{50}{\um} in thickness.

INS spectra were measured on the IN8 high-flux thermal neutron triple-axis spectrometer \cite{Piovano2020} at Institut Laue Langevin (ILL)~\cite{ExperimentIN8}. The initial and final neutron energies were selected using a doubly focused Cu (200) monochromator and analyzer, which resulted in an energy resolution of \SI{0.62}{\meV} as measured at the elastic peak. A pyrolytic graphite filter was placed in front of the analyzer to suppress higher-order neutrons. Measurements were performed at a fixed scattered wave vector $k_f = \SI{2.662}{\per\angstrom}$. Due to the reduced size of the crystal, no collimation could be installed without excessive loss in intensity. A preliminary test experiment was also previously carried out on the triple-axis spectrometer IN22 at the ILL with similar but more limited results~\cite{ExperimentIN22}. 

IXS measurements were performed on the ID28 beam line at the European Synchrotron Radiation Facility (ESRF), as described in Refs. \cite{Krisch2006, Girard2019}. For this experiment, the (999) Bragg reflection of a Si crystal was used to analyse the scattered beam, giving rise to an energy resolution $\Delta E = 3.0 \pm 0.2$ \SI{}{\milli\electronvolt}. 



\subsection{R-point data from INS}

\begin{figure}[ht]
    \centering
    \includegraphics[width=2.8 in]{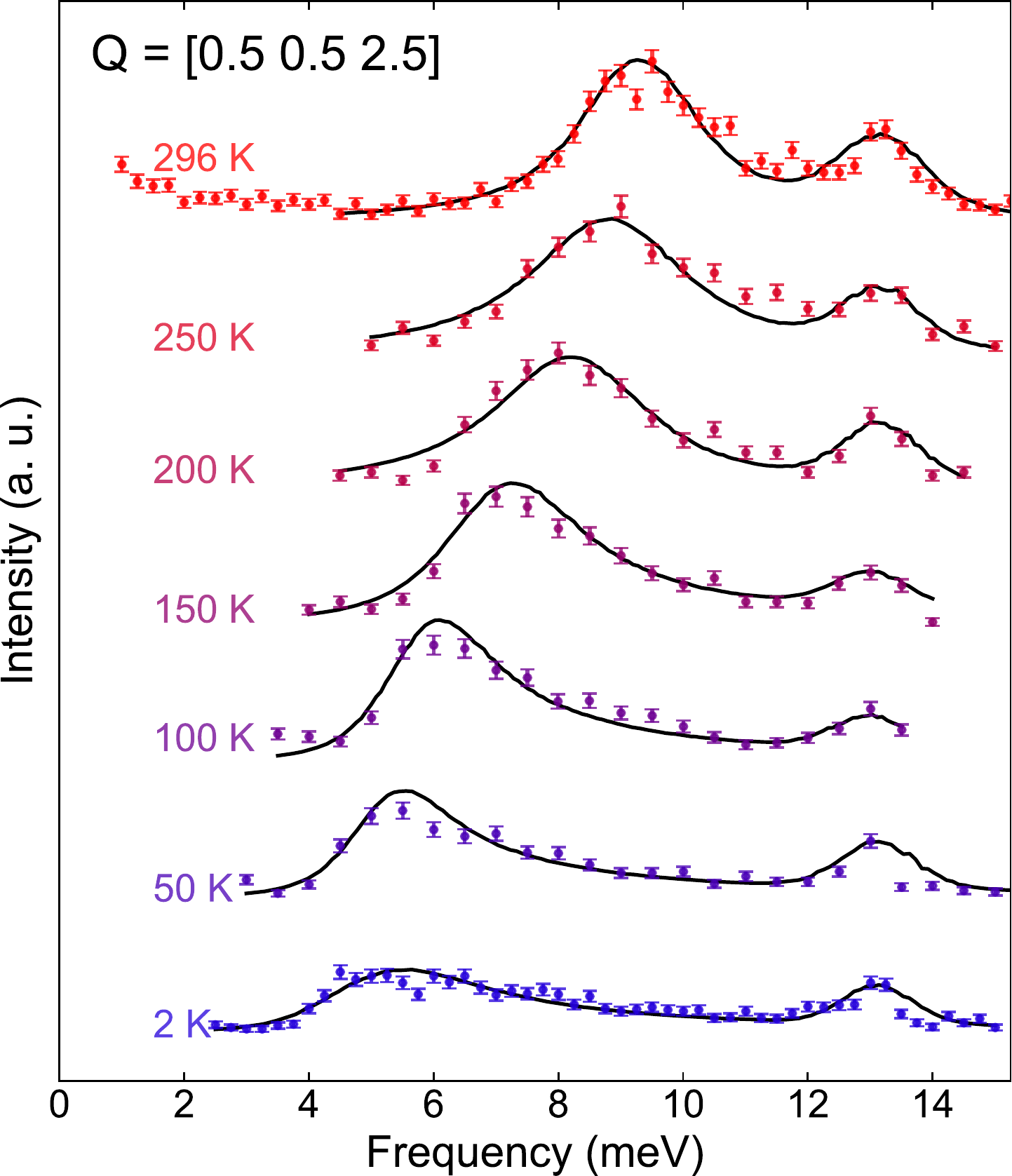}
    \caption{Inelastic neutron scattering (INS) spectra measured at the R-point (0.5 0.5 2.5) between \SI{296}{\kelvin} and \SI{2}{\kelvin}, showcasing the softening of the lowest-energy mode.
    }
    \label{fig:INS_R}
\end{figure}

Phonon spectra at the R-point (0.5 0.5 2.5) were measured by INS at different temperatures and the results are presented in Fig.~\ref{fig:INS_R}. 
The spectra display two peaks at every temperature between \SI{296}{\kelvin} to \SI{2}{\kelvin}. The low-frequency peak shows a clear softening accompanied by an increase of the asymmetry of the peak, while the higher-frequency peak is essentially temperature independent in the considered temperature interval.

By comparing with DFT calculations and the calculated phonon dispersions in Fig.~\ref{fig:harmonic_dispersion} the low-frequency peak can be attributed to the antiphase oxygen tilt oscillations, the R-tilt mode, while the higher-frequency peak corresponds to the acoustic branch and is connected to barium displacements, the R-acoustic mode. We also checked that the R-tilt mode vanishes at the R point (1.5 1.5 1.5) while the R-acoustic mode is still present, which is consistent with the neutron selection rules expected for modes of $R_4^+$ et $R_5^+$ symmetry, respectively~\cite{ExperimentIN8}.

INS spectra of BaZrO$_3$ are proportional to the neutron coherent cross-section: $(\Bar{b})^2 \frac{k_f}{k_i} S(\boldsymbol{Q}, E)$, where $b$ is the nuclear scattering length, $k_i$ (respectively $k_f$) the wave vector of the incident (respectively scattered) neutron beam and $S(\boldsymbol{Q}, E)$ the neutron scattering function. For oxygen and barium, the coherent scattering lengths are quite similar ($b_\textrm{O}=5.803$ fm and $b_\textrm{Ba}=5.07$ fm) while zirconium has a higher one ($b_\textrm{Zr}=7.16$ fm) \cite{Neutron-scattering}. Hence, for given $k_i$ and $k_f$ and with the proper selection rules, neutron intensity should be of the same order of magnitude for the oxygen-dependent R-tilt and the barium-dependent R-acoustic modes. 


Besides the pronounced softening, the R-tilt mode shows a strong asymmetric shape. We assume that this is due to the strong dispersion of the phonon branch close to the R-point. The increase of the phonon frequency away from the R-point implies that the measured peak may obtain an asymmetric shape towards higher frequencies.
Anharmonicity also contributes to the broadening of the R-tilt mode but that broadening is more symmetric in frequency  and increases with temperature ({\it cf.} \autoref{fig:dynasor_spectra}).
Nanodomains, proposed in Ref~\cite{levinNanoscalecorrelatedOctahedralRotations2021}, could arguably affect the mode damping and the spectrum in general, although in the present state, our data do not really provide evidence for or against the presence of these domains. 

The neutron scattering data were then fitted with a sum of damped harmonic oscillators using the Takin \cite{Weber2016, Weber2017, Weber2021} software, taking into account the instrumental resolution (see Fig.~S4 in Supplemental Material \cite{SI:this}). In order to account for the asymmetric shape of the R-tilt mode, a first fitting was attempted by assuming a parabolic dispersion. This however did not allow us to obtain a satisfactory fit. 
We therefore introduced a more ad hoc mathematical asymmetry to the lineshape. This allows us to extract values for the mode energy, but does not allow to deconvolute fully the spectrum and get access to the intrinsic widths. With this data treatment, we observe a softening of the R-tilt mode from \SI{9.4}{\milli\electronvolt} at \SI{296}{\kelvin} to \SI{5.6}{\milli\electronvolt} at \SI{2}{\kelvin}. The R-acoustic mode stays essentially constant around an average value of \SI{13.0}{\milli\electronvolt}. 

The strong temperature dependence of the R-tilt mode contrasts with the inelastic scattering data previously reported by Perrichon et al. \cite{perrichonUnravelingGroundStateStructure2020} on a \emph{powder} sample. At low temperatures, \SI{5}{\kelvin}, they obtained the frequency \SI{5.8}{\milli\electronvolt}, in very good agreement with our result. However, essentially no temperature dependence could be seen in the interval from \SI{5}{\kelvin} to \SI{500}{\kelvin}. The present single crystal study allows us to observe and quantify the temperature dependence more accurately.

\subsection{R-point data from IXS}

\begin{figure}[htpb]
    \centering
    \includegraphics[width=2.8 in]{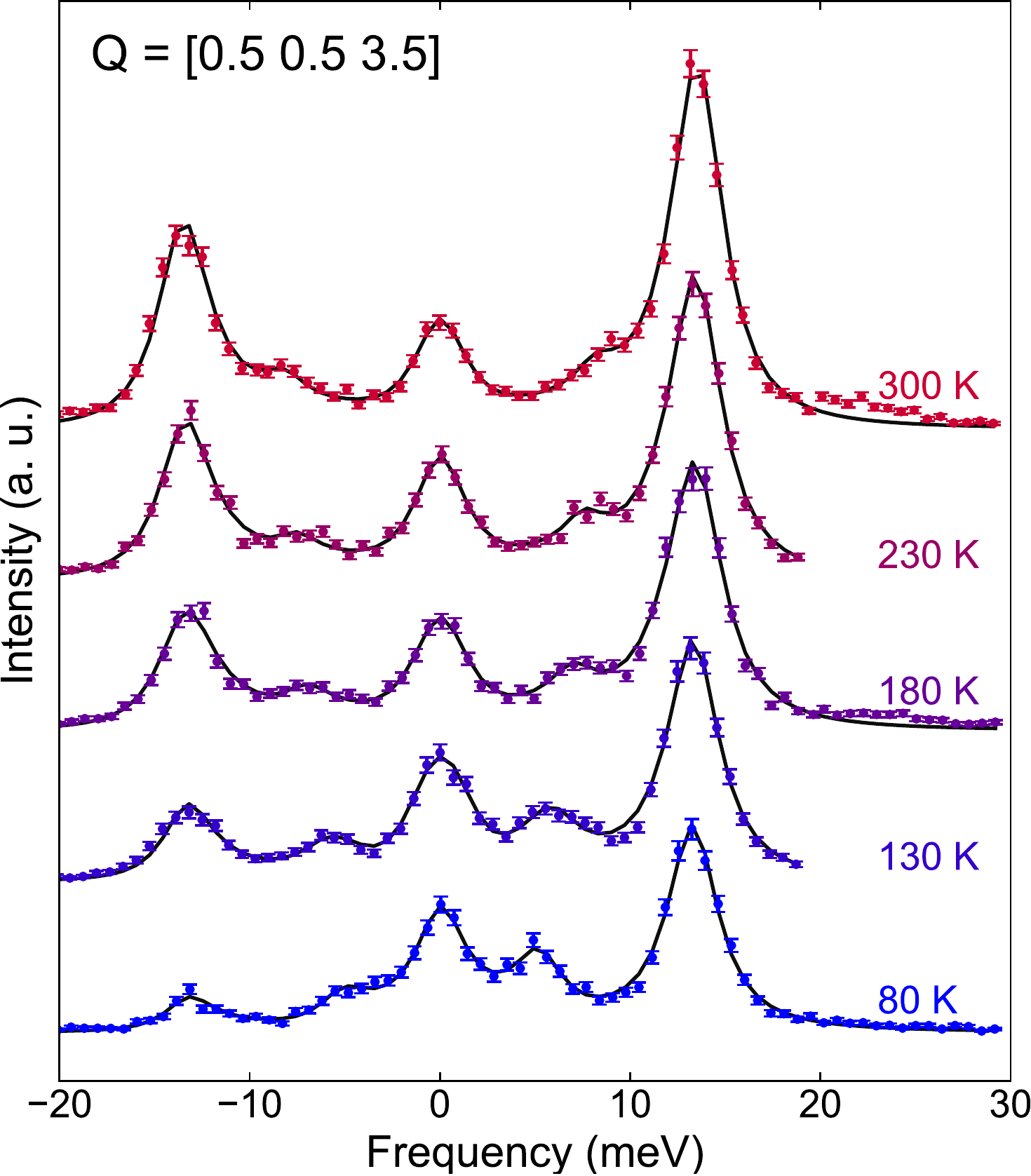}
    \caption{Inelastic x-ray scattering (IXS) spectra measured at the R-point (0.5 0.5 3.5) between \SI{300}{\kelvin} and \SI{80}{\kelvin}. Both Stokes and anti-Stokes peaks are represented. 
    We observe the softening of the R-tilt mode. The R-acoustic mode is barely temperature-dependent.}
    \label{fig:IXS_R}
\end{figure}

The phonon spectra at the R-point were also measured by IXS at different temperatures and the result is presented in Fig.~\ref{fig:IXS_R}. 
At each temperature between \SI{300}{\kelvin} to \SI{80}{\kelvin} the spectrum displays a central quasi-elastic peak and two peaks respectively linked to the R-tilt mode and the R-acoustic mode, which can be both seen at positive and negative frequencies, due to Stokes and anti-Stokes scattering processes. Here, in contrast to the INS measurement, the R-tilt mode involving oxygen motion appears much weaker than the acoustic mode that is dominated by barium displacements. The IXS spectra were fitted by the sum of two damped harmonic oscillators and a Lorentzian for the pseudo-elastic contribution. The R-tilt mode frequencies softens from \SI{8.3}{\milli\electronvolt} at \SI{300}{\kelvin} to \SI{5.2}{\milli\electronvolt} at \SI{80}{\kelvin}. The R-acoustic mode stays essentially constant around an average value \SI{13.4}{\milli\electronvolt}. The widths of the modes are in all cases resolution limited and do not show any physically insightful trend. 



\subsection{Dispersion data from INS and IXS}

\begin{figure}[htpb]
    \centering
    \includegraphics[width=3.2 in]{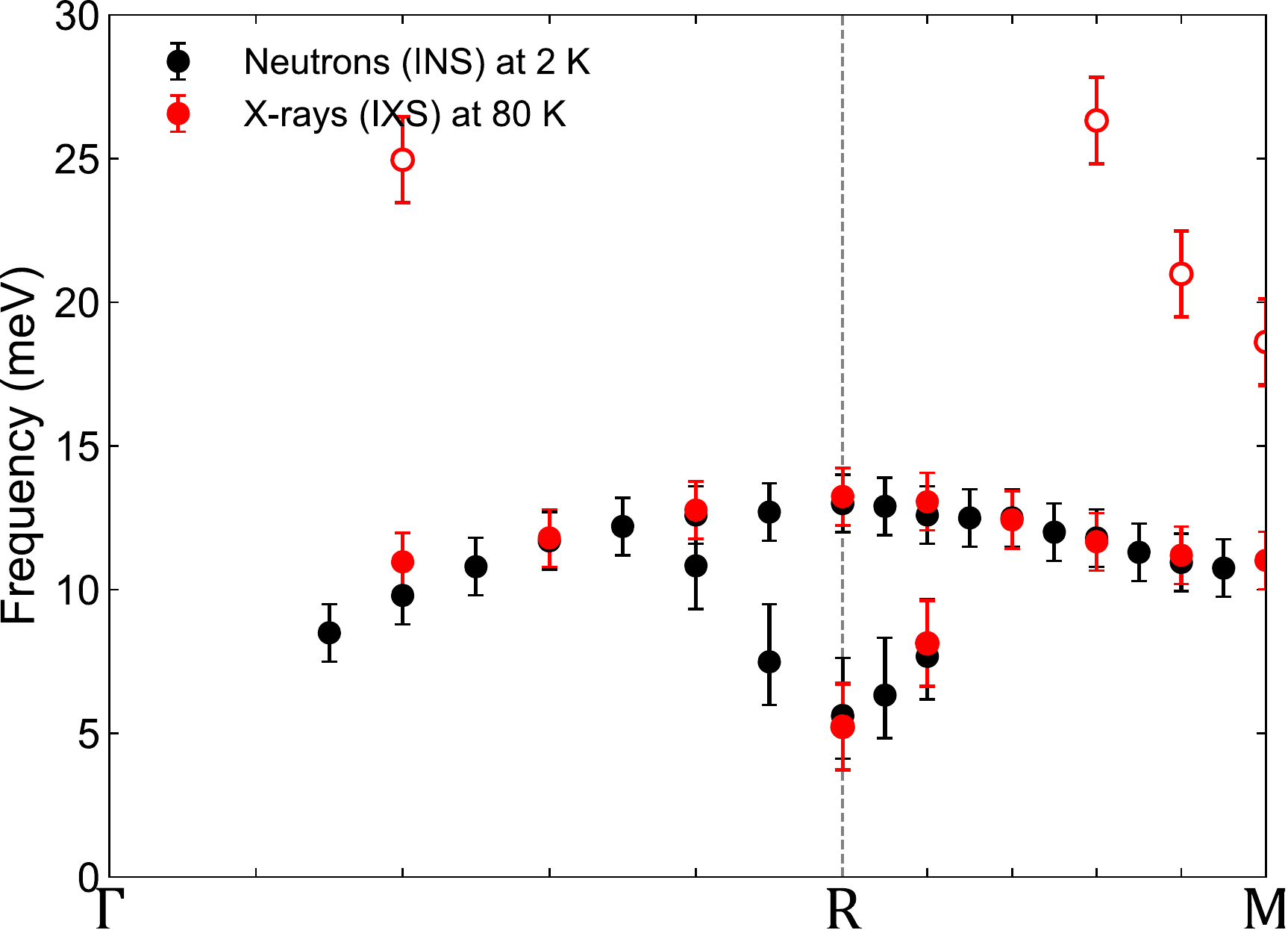}
    \caption{Measured dispersion curves along the $\Gamma$--R--M path, measured at \SI{2}{\kelvin} by inelastic neutron scattering (INS) and at \SI{80}{\kelvin} by inelastic x-ray scattering (IXS). For experimental data, error bars are the width of the fitted peak (INS) or the instrumental resolution (IXS). Open symbols indicate data points that we could not identify by comparison to phonon calculations.
 } 
    \label{fig:Exp_dispersion}
\end{figure}

To reconstruct phonon dispersions, we have also measured spectra away from the R-point, in the directions  R(0.5 0.5 2.5)--M(0.5 0.5 2) and R(0.5 0.5 2.5)--$\Gamma$(1 1 2) at \SI{2}{\kelvin} for INS. For IXS, we measured in the directions  R(0.5 0.5 3.5)--M(0.5 0.5 4) and R(0.5 0.5 3.5)--$\Gamma$(1 1 4) at \SI{80}{\kelvin}. When both R-tilt and R-acoustic modes were resolved, INS spectra were fitted as previously. However, further from R, it becomes very difficult to resolve two peaks as the R-tilt mode broadens and crosses the R-acoustic mode. Then, spectra were fitted by a single-damped harmonic oscillator to extract the frequency of the R-acoustic mode. For the IXS experiments, spectra were fitted as previously by a sum of damped harmonic oscillators for the R-tilt mode and R-acoustic mode, and by a Lorentzian lineshape for the pseudo-elastic peak centered around \SI{0}{\milli\electronvolt}. In some spectra, higher frequencies features appear above \SI{20}{\milli\electronvolt} and up to the edge of the frequency range ($\approx$ \SI{30}{\milli\electronvolt}). When these features are present, they have been fitted by additional damped harmonic oscillators. 

Frequencies extracted from the fits of INS and IXS spectra are represented in Fig.~\ref{fig:Exp_dispersion}. The spectra and their fits are given in Figs S5 and S6 in Supplemental Material \cite{SI:this}. We can first observe that the frequency of the R-acoustic mode is very consistent between the INS and IXS measurements, the difference in temperature for the two measurements being here negligible. The R-acoustic mode reaches its maximum frequency of around \SI{13}{\milli\electronvolt} at the R-point. This frequency then decreases towards $\Gamma$ and tends towards a value of around \SI{11}{\milli\electronvolt} at the M-point.

The R-point soft-mode is at its lowest energy at R (in INS $\approx$ \SI{5.6}{\milli\electronvolt}, in IXS $\approx$ \SI{5.2}{\milli\electronvolt})  and rises quickly in energy as we probe further from R. While the spectra are particularly unambiguous at the R point itself, the picture becomes much less clear as we move away from the zone boundary. Along the R-M direction, we observe a single branch attributed to oxygen, whereas phonon band calculations and symmetry considerations predict a splitting into two oxygen branches, see Fig.~\ref{fig:harmonic_dispersion}. The tilt modes at M also could not be identified with certainty. This is because the lowest tilt mode at M (with M$_3^+$ symmetry) as well as the branch (T$_4$) connecting that mode to the tilt mode at R are indeed forbidden by symmetry for our measurement configuration with $Q$ from (0.5 0.5 2.5) to (0.5 0.5 2.0). Further experiments in a less symmetric configuration assisted with accurate calculations of scattering cross sections will be needed to clarify the modes at M and the dispersion along the R-M direction. 

In addition, in the IXS data we observe extra modes at higher frequencies (above \SI{20}{\meV}, shown in open symbols on Fig.~\ref{fig:Exp_dispersion}) both in the middle of the R-$\Gamma$ direction and in the R-M direction. The energies of the modes on the R-M direction agree very well with the Zr-dominated (green) phonon branch seen in Fig.~\ref{fig:harmonic_dispersion}, and the high intensity of the IXS peaks also points to a mode involving a heavy atom. The mode along $\Gamma-R$ on the other hand is much weaker (see Fig.~S6 in Supplemental Material \cite{SI:this}) and difficult to assign conclusively. This mode might be of spurious origin and further studies will be necessary to clarify its assignment. This is inconsequential for our low-frequency study of the R-tilt mode and acoustic mode.

\section{Theoretical results and comparison}
\label{sec:theory_results}

\subsection{Dynamical structure factor}

The \gls{fcp} potential can be used to perform \gls{md} simulations. The full anharmonicity in \eq{eq:fcp_expansion} is taken into account but the motion is restricted to classical dynamics. 

We consider an 8x8x8 supercell ($N$=2560 atoms) with periodic boundary conditions and the \gls{md} simulations are carried out using the \textsc{GPUMD}~\cite{fanEfficientMolecularDynamics2017} package with the \gls{fcp} \cite{brorssonEfficientCalculationLattice2022}.
The system is equilibrated in the NVT ensemble using the Langevin thermostat and the positions $\boldsymbol{r}_i(t) = \mathbf{R}_i^0 + \mathbf{u}_i(t)$ are then sampled in the NVE ensemble using the velocity Verlet algorithm. The total simulation time is set to $\SI{2}{\nano\second}$ with a time step of $\SI{1}{\femto\second}$.

The intermediate scattering function
\begin{align}
    F(\boldsymbol{q},t) = \frac{1}{N} \sum_i^N \sum_j^N \left < b_i b_j \mathrm{exp} \left [ i \boldsymbol{q} \cdot ( \boldsymbol{r}_i(t)-\boldsymbol{r}_j(0)) \right ] \right >
\end{align}
is determined at the R point, for the wavevector $\boldsymbol{q}=(2\pi/a)[0.5,0.5,2.5]$, as a time average 
$\langle \cdots \rangle
$
using \dynasor{} \cite{FraSlaErhWah2021}. The nuclear scattering lengths $b_i$ are given in Sec.~\ref{sec:exp_results}. The dynamic structure factor
\begin{align}
    S(\boldsymbol{q},\omega) = \int_{-\infty}^\infty F(\boldsymbol{q},t)
    \mathrm{exp} \left ( -i \omega t \right ) \mathrm{dt}
\end{align}
is then obtained by a Fourier transformation and $S(\boldsymbol{q},\omega)$ is averaged over 50 independent \gls{md} simulations with different initial velocities.

\begin{figure}[ht]
\centering
\includegraphics{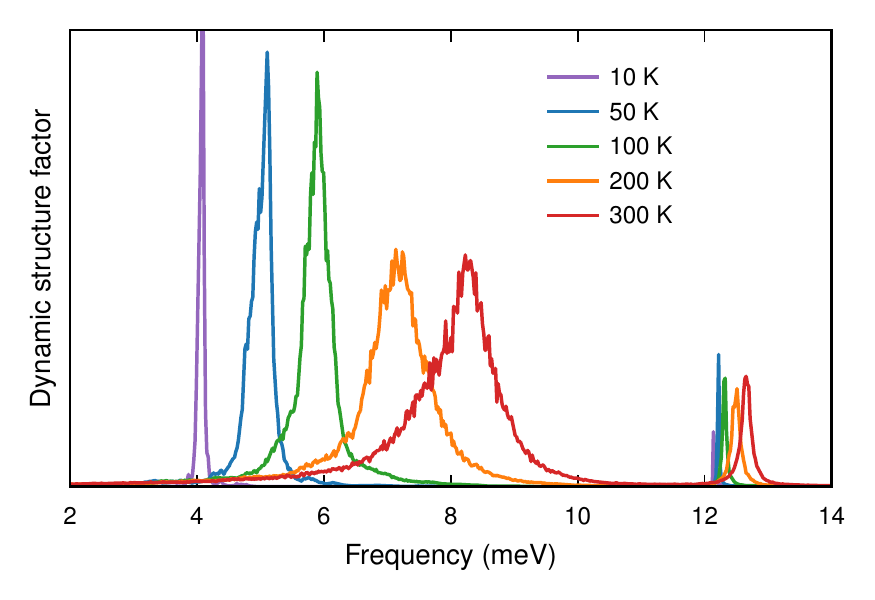}
\caption{The temperature dependency of the dynamical structure factor $S(\boldsymbol{q},\omega)$ at the R point ($\boldsymbol{q}=(2\pi/a)[0.5,0.5,2.5]$) from MD simulations using the force constant potential (\gls{fcp}) model based on PBE.}
\label{fig:dynasor_spectra}
\end{figure}

The resulting dynamical structure factor, as function of temperature, is shown in \autoref{fig:dynasor_spectra}.
Two set of peaks are seen.
One at lower frequencies, from 4 to \SI{9}{meV}, and one at higher frequencies, between 12 and \SI{13}{meV}.
The lower frequency mode corresponds to oxygen motion, the R-tilt mode. It shows a strong temperature dependence, and the peak broadens as temperature increases.
The peaks at higher frequencies correspond to motion of barium, the R-acoustic mode. A weak temperature dependence is observed, indicating that this vibrational motion of barium is associated with a weak anharmonicity.
In order to extract the frequencies, we fit the spectra to a damped harmonic oscillator model (see Supplemental Material \cite{SI:this}). The obtained theoretical frequencies are compared with our experimental data points in \autoref{fig:rmod_frequency}.

\subsection{Dispersion relation}

Next, we consider the \gls{scp} method. The model is based on a 6x6x6 supercell, and we use M=80 structures in each iteration.
The minimization in \eq{eq:scph_minimization} is solved by \gls{ols}
and the iteration in \eq{eq:scph} is stopped once the free energy for the effective harmonic system is smaller than $\SI{0.001}{meV}$ per atom for two successive iterations.

\begin{figure}[ht]
\centering
\includegraphics{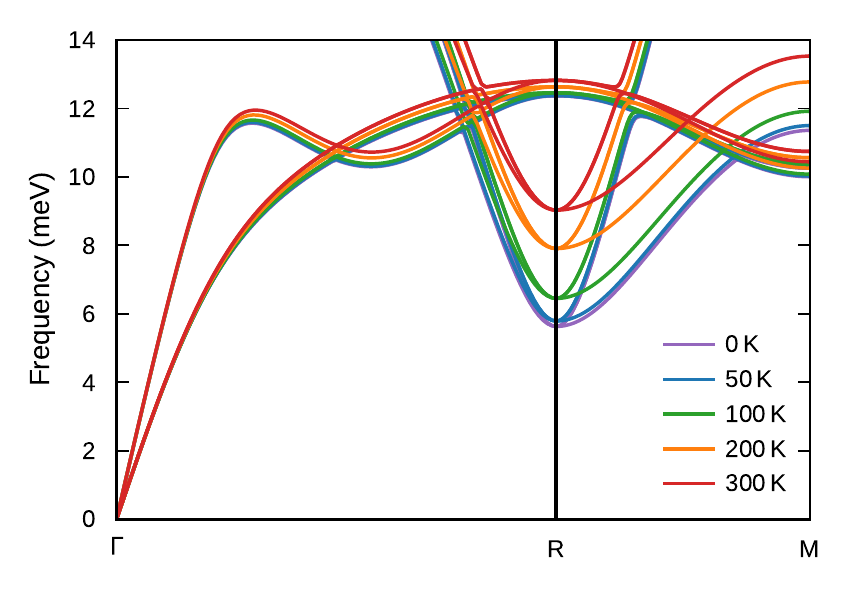}
\caption{
    The phonon dispersion at different temperatures obtained from the self consistent phonon (\gls{scp}) model based on PBE.
}
\label{fig:ehm_dispersion}
\end{figure}

In \autoref{fig:ehm_dispersion} part of the resulting phonon dispersion for five different temperatures is shown. The oxygen related R-tilt mode is clearly temperature dependent, and the barium related R-acoustic mode shows only a weak temperature dependence.

At the R-point we observe a softening of the R-tilt mode from
\SI{9.0}{\milli\electronvolt} at \SI{300}{\kelvin} to \SI{5.6}{\milli\electronvolt} at \SI{0}{\kelvin}. 
Also at the M-point the softening is substantial, from \SI{13.5}{\milli\electronvolt} to 
\SI{11.4}{\milli\electronvolt} 
in the same temperature range.
The R-acoustic mode is nearly constant with temperature. At the R-point the frequency decreases slightly, from  
\SI{12.8}{\milli\electronvolt} to 
\SI{12.4}{\milli\electronvolt} 
in the temperature range \SI{300}{\kelvin} to \SI{0}{\kelvin}. 
The obtained theoretical frequencies are compared with our experimental data points in \autoref{fig:rmod_frequency} and \autoref{fig:rmod_dispersion}

\subsection{Comparison between theory and experiments}

\begin{figure}[ht]
\centering
\includegraphics{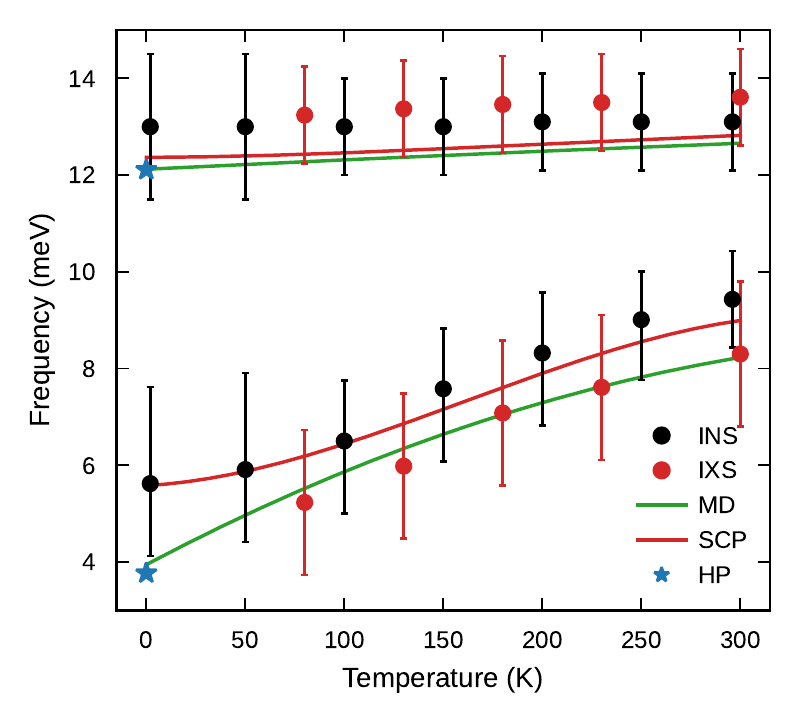}
\caption{
    The temperature dependency for the R-tilt and R-acoustic modes, both from experiments (INS and IXS) and theory (SCP and MD). HP is the harmonic approximation, which is temperature independent. The theoretical data are based on PBE. For experimental data, error bars are the width of the fitted peak (INS) or the instrumental resolution (IXS).
}
\label{fig:rmod_frequency}
\end{figure}

Our key result is the temperature dependence of the antiferrodistortive phonon mode, the R-tilt mode. In \autoref{fig:rmod_frequency} we compare our experimental and theoretical results for both the R-tilt mode and the R-acoustic mode.

The R-tilt mode shows a strong temperature dependence. From the INS data we obtain a softening from \SI{9.4}{\milli\electronvolt} at \SI{296}{\kelvin} to \SI{5.6}{\milli\electronvolt} at \SI{2}{\kelvin}, while the IXS data decrease from \SI{8.3}{\milli\electronvolt} at \SI{300}{\kelvin} to \SI{5.2}{\milli\electronvolt} at \SI{80}{\kelvin}. 
The INS frequencies are therefore consistently higher by about \SI{1}{\milli\electronvolt}, compared with the IXS frequencies.
The two data set agree within experimental uncertainties, but 
we consider the INS data to be more reliable since the R-tilt mode is weak in the IXS spectra. 
Our theoretical results, the data based on the self-consistent phonon (SCP) method and the result obtained using the dynamic simulation technique (MD), agree nicely with the experimental data.

Theory predicts that below about \SI{100}{K} the quantum fluctuations of the atomic motion become important in sampling the configuration space (cf. \autoref{fig:rmod_frequency_PBE_PBEsol}). It implies that the change of the frequency, the slope, decreases when approaching \SI{0}{K} from above. This is seen in the experimental INS data, while the IXS measurements do not reach sufficiently low temperatures to observe this effect. The quantum fluctuations are included in the SCP method and the corresponding data show a clear decrease of the slope. In the MD method, the dynamics is treated classically (does not contain the quantum fluctuations) and the corresponding data show an increase of the slope when approaching 0 K from above, in contrast to the SCP data.

At higher temperatures, above \SI{100}{K}, the R-tilt mode is well described by classical sampling and thus the difference between \gls{md} and \gls{scp} seen in \autoref{fig:rmod_frequency} comes from the differences between these two computational techniques.
In the MD technique all anharmonicities within the given FCP are included while in the SCP method the configuration space is sampled by independent normal modes, based on an effective harmonic model which has been computed using the same FCP as in MD.
It has been argued that, using non-interacting normal modes, could potentially lead to sampling of higher-energy structures by the SCP method.
This would then result in higher effective vibrational frequencies \cite{metsanurkSamplingdependentSystematicErrors2019}.
Therefore, we expect the MD method to be more accurate compared with the SCP method at higher temperatures.

For comparison, the results using the effective harmonic model introduced by Hellman {\it et al.} \cite{hellmanLatticeDynamicsAnharmonic2011} have also been computed.
This model is based on samples from a classical MD trajectory and thus lacks the quantum fluctuations of the atomic motion.
In the present case, it gives very similar results as the \gls{md} model for the temperature dependence of the R-tilt mode. For more details, see Fig.~S8 in Supplemental Material \cite{SI:this}. 

In \autoref{fig:rmod_frequency} we also show the temperature dependence of the R-acoustic mode, located at about \SI{13}{\milli\electronvolt}. It shows a very weak temperature dependence. The IXS data decreases with about \SI{0.4}{\milli\electronvolt} when reducing the temperature from \SI{300}{\kelvin} down to \SI{80}{\kelvin}, while the INS data are essentially temperature independent. A small decrease is also obtained using the SCP and MD methods, about \SI{0.5}{\milli\electronvolt}. In this case the two theoretical methods give very similar results for the frequencies. The theoretical data are based on PBE and are about \SI{1}{\milli\electronvolt} lower than the experimental result. However, it is known that the more accurate hybrid functionals HSE and CX0p give about \SI{1}{\milli\electronvolt} higher values for the R-acoustic mode, compared with PBE \cite{granhedBaZrOStabilityPressure2020}. 

\begin{figure}[ht]
\centering
\includegraphics{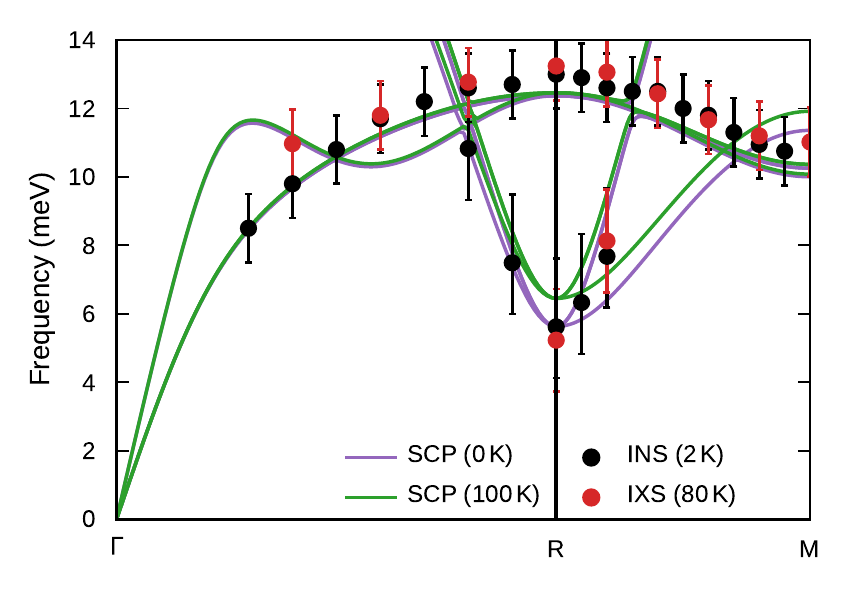}
\caption{
       The dispersion curves for the R-tilt and R-acoustic modes, both from experiments (INS and IXS) and theory (SCP), and at two different temperatures, about 0 K and 100 K. The theoretical data are based on PBE. For experimental data, error bars are the width of the fitted peak (INS) or the instrumental resolution (IXS).
}
\label{fig:rmod_dispersion}
\end{figure}

Finally, we compare the experimental data for the dispersion of the two R-modes with the theoretical predictions based on the SCP method in \autoref{fig:rmod_dispersion}. The INS data for the R-tilt mode compare very well with the theoretical data, while in the IXS measurement only one extra data point could be resolved. In the direction R--M theory predicts two branches but both the INS and the IXS measurements could only resolve one branch, as discussed earlier. The R-acoustic mode is also very well reproduced by both the INS and IXS data. A systematic offset is visible, 
but as discussed above this is most likely due to the use of PBE in the theoretical calculations.


\section{Theoretical results - atomic displacements}
\label{sec:theory_results_atomic_displacemnts}

\subsection{Mean squared displacements}

\begin{figure}[!ht]
\centering
\includegraphics{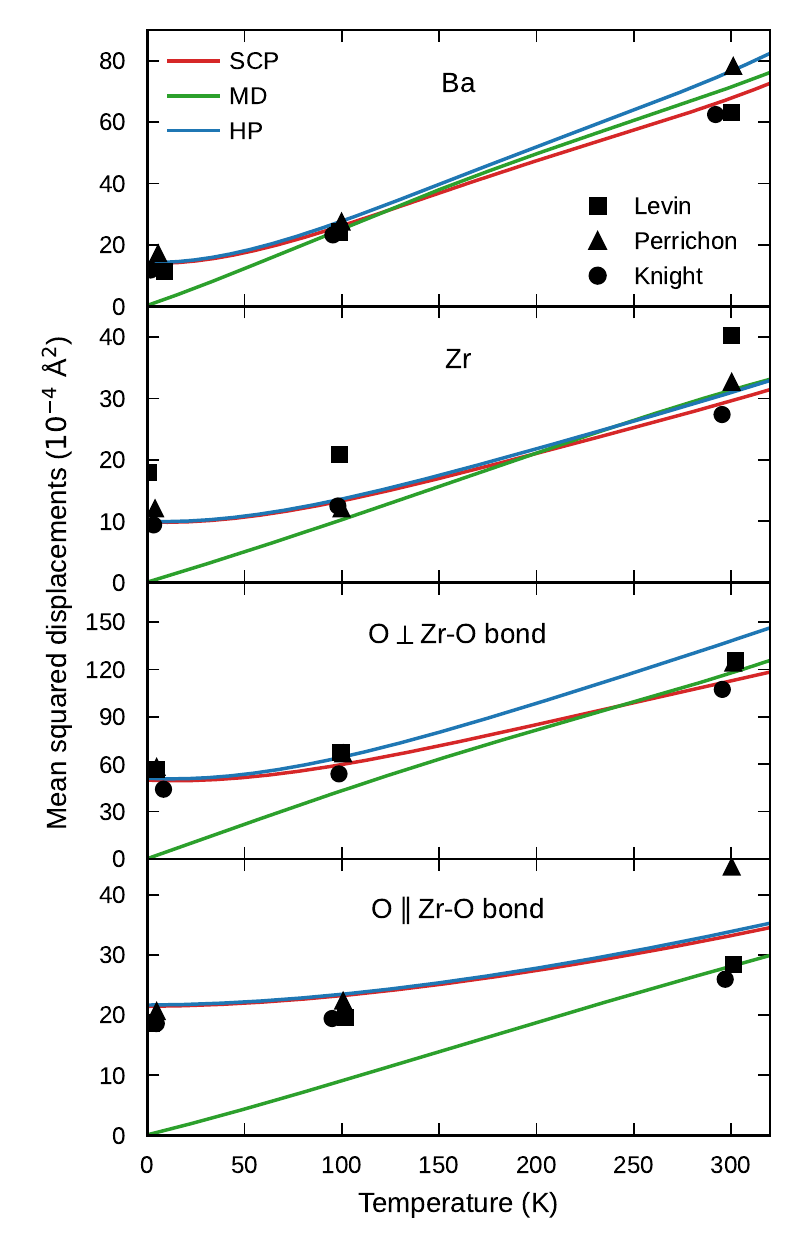}
\caption{
    Mean squared displacements from \gls{md}, \gls{scp} and \gls{hp}, based on PBE. Experimental data taken from  \citet{levinNanoscalecorrelatedOctahedralRotations2021}, \citet{perrichonUnravelingGroundStateStructure2020} and \citet{knightLowtemperatureThermophysicalCrystallographic2020}.
}
\label{fig:mean-squared-displacement}
\end{figure}

The mean squared displacement (MSD) $\langle (u_{i}^{\alpha})^2 \rangle$ can be obtained within the HP and SCP models using \autoref{eq:u2_harm}.
It can also be obtained directly from MD simulations making a time-average along the MD trajectory
\begin{equation}
   \langle (u_{i}^{\alpha})^2 \rangle = 
    \langle (u_{i}^{\alpha}(t))^2 \rangle_{\textrm{time}}
   \label{eq:u2_md}\ ,
\end{equation}
where $\mathbf{u}_i(t)$ is the displacement of atom $i$ from its equilibrium position $\mathbf{R}_i^0$, {\it i.e.} $\boldsymbol{r}_i(t) = \mathbf{R}_i^0 + \mathbf{u}_i(t)$.
The obtained results for the \glspl{msd} are shown as function of temperature in \autoref{fig:mean-squared-displacement}. In the same figure we also show experimental data for BZO from Refs
\cite{perrichonUnravelingGroundStateStructure2020,knightLowtemperatureThermophysicalCrystallographic2020,levinNanoscalecorrelatedOctahedralRotations2021}.

We notice that for barium and zirconium all directions are symmetrically equivalent. Barium has a larger displacement compared with zirconium due to its considerably weaker interaction with its surroundings. On the other hand, the displacement for oxygen is strongly anisotropic, with larger displacements perpendicular to the Zr-O bond.

Consider first the \gls{hp} model. At zero Kelvin the MSD is given by its zero point motion value. It increases with temperature and at high temperatures it approaches its classical behavior, i.e. linear in temperature.

In the \gls{scp} model anharmonicity is included. There is a clear difference between \gls{scp} and \gls{hp} at higher temperatures for the oxygen motion perpendicular to the Zr-O bond. The \gls{fcp} based \gls{scp} potential energy landscape is stiffer than the harmonic (\gls{hp}) one (cf. \autoref{fig:model_validation_Rmode_pes}) and thus leading to smaller displacements. The displacements for barium show a similar effect but less pronounced.

In the \gls{md} model the motion is treated classically.
Therefore, at low temperatures the displacements approaches zero.
This is seen in \autoref{fig:mean-squared-displacement}.
Furthermore, if the system is harmonic the MSD increases linearly with temperature.
This is obtained for the zirconium motion and the motion of oxygen parallel to the Zr-O bond.
However, for the motion of oxygen perpendicular to the Zr-O bond the slope of the \gls{msd} decreases when the temperature increases, due to the motion being anharmonic.
The same behavior is observed for the barium motion but less pronounced.

The difference between the \gls{md} and the \gls{scp} models is that the latter includes the quantum fluctuations of the atomic motion, while \gls{md} does not take those into account. On the other hand, \gls{md} includes the full anharmonicity in \autoref{eq:fcp_expansion}, while \gls{scp} does that in an approximate way.
From \autoref{fig:mean-squared-displacement} we note that \gls{md} and \gls{scp} agree quite well at \SI{300}{K} for all atomic motions except for the oxygen motion parallel to the Zr-O bond. For this high frequency motion, quantum effects are sizeable even at \SI{300}{K}.

We can also compare with available experimental data.
In \autoref{fig:mean-squared-displacement} data at \SI{5}{K}, \SI{100}{K} and \SI{300}{K} from Perrichon {\it et al.} \cite{perrichonUnravelingGroundStateStructure2020}, Knight \cite{knightLowtemperatureThermophysicalCrystallographic2020} and Levin {\it et al.} \cite{levinNanoscalecorrelatedOctahedralRotations2021} are added. The mean square static disorder has been subtracted from the raw data by Knight \cite{knightLowtemperatureThermophysicalCrystallographic2020}. We notice that theory and experiments compare quite well. However, the data by Levin {\it el al.} \cite{levinNanoscalecorrelatedOctahedralRotations2021} for Zr are systematically larger and the \SI{300}{K} data point for oxygen parallel to the Zr-O bond
by Perrichon {\it et al.} \cite{perrichonUnravelingGroundStateStructure2020} deviates significantly from the rest of the experimental data.

\subsection{Mean squared relative displacements}

The fluctuations in the distance between two atoms in a crystal can be measured using extended X-ray absorption fine structure (EXAFS) spectroscopy \cite{fornasiniEXAFSDebyeWallerFactor2015}. This technique gives a more local information about the fluctuations of the atomic positions compared with the mean squared displacement (MSD), measured by X-ray and neutron scattering.

The R-tilt mode with its essentially rigid antiphase tilts of the ZrO$_6$ octahedra should be visible as large fluctuations of the nearest-neighbor Ba-O atomic distance. Indeed, Lebedev and Sluchinskaya \cite{lebedevStructuralInstabilityBaZrO2013} found large values of the Debye Waller factor for the Ba-O atomic pairs using EXAFS at \SI{300}{K}. Usually the Debye-Waller factor in EXAFS monotonically increases with increasing inter-atomic distances, but Lebedev and Sluchinskaya \cite{lebedevStructuralInstabilityBaZrO2013} found
a significantly larger value for the first shell (Ba-O) compared with the second (Ba-Zr) and third (Ba-Ba) shells (cf. \autoref{tab:MSRD}). They attributed this to the appearance of a structural instability and speculated in the possible formation of a structural glass state in BZO as the temperature is lowered.

Granhed {\it et al.} \cite{granhedBaZrOStabilityPressure2020} showed, using DFT calculations, that the low frequency of the R-tilt mode gives rise to an "anomalously" large value of the Debye-Waller factor for the Ba-O atomic pairs. The study by Granhed {\it et al.} \cite{granhedBaZrOStabilityPressure2020} was based on the harmonic approximation for the lattice vibrations and several exchange-correlation functionals within DFT were used.


Here, we consider the effect of the anharmonicity which was neglected in Ref. \cite{granhedBaZrOStabilityPressure2020}.
The parallel \gls{msrd} between neighboring atoms $a$ and $b$ can be obtained from the phonon eigenvectors and frequencies according to \cite{fornasiniEXAFSDebyeWallerFactor2015},
\begin{equation} \label{eq:msrd2}
   \langle \Delta u^2_\parallel \rangle_{ab} = 
   \frac{1}{N_q}\sum_{{\boldsymbol{q}},\nu}
   \frac{\hbar}{2\mu_{ab}\omega_{{\boldsymbol{q}},\nu}}
   \mid Y_{\boldsymbol{q},\nu}^{ab} \mid^2
   \coth{(\hbar\omega_{{\boldsymbol{q}},\nu}/2k_{\text B}T)}
\end{equation}
where 
\begin{equation} \label{eq:Yab}
    Y_{\boldsymbol{q},\nu}^{ab} =
    \left[\left(\frac{\mu_{ab}}{m_b}\right)^{1/2}\mathbf{e}^b_{\boldsymbol{q},\nu}e^{i\boldsymbol{q}\cdot\mathbf{R}_{ab}^0} - 
          \left(\frac{\mu_{ab}}{m_a}\right)^{1/2}\mathbf{e}^a_{\boldsymbol{q},\nu}\right] \cdot \hat{\mathbf{R}}_{ab}^0
\end{equation}
and $\mu_{ab} = m_am_b/(m_a+m_b)$ is the reduced mass, $\mathbf{R}_{ab}^0$ is the vector connecting the equilibrium positions of neighboring atoms $a$ and $b$ and $\hat{\mathbf{R}}_{ab}^0$ is the corresponding vector of unit length. Eqs. (\ref{eq:msrd2}) and (\ref{eq:Yab}) can be used to obtain the parallel \gls{msrd} within the \gls{hp} and \gls{scp} approximations.
The parallel \gls{msrd} can also be directly computed from \gls{md} simulations by making a time average of the expression
\begin{equation} \label{eq:msrdmd}
   \langle \Delta u^2_\parallel \rangle_{ab} =
   \langle\  \mid
   \left( \mathbf{u}_a(t) - \mathbf{u}_b(t) \right) \cdot \hat{\mathbf{R}}_{ab}^0 
   \mid^2 \rangle_{\textrm{time}}
\end{equation}
along the MD trajectory \cite{fornasiniEXAFSDebyeWallerFactor2015}.

Our resulting parallel MSRDs are presented in \autoref{tab:MSRD} using PBE. 
In the harmonic approximation (HP) we indeed get a large value for the nearest-neighbor Ba-O atomic distance at \SI{300}{K}, considerably larger than the experimental value.
That was also noticed in Ref. \cite{granhedBaZrOStabilityPressure2020} and explained by the underestimation of vibrational frequencies in PBE. 
If we add anharmonicity and evaluate the \gls{msrd} within the \gls{scp} and \gls{md} approximations the value is reduced but still larger compared with experiments. 
This reduction is due to the stiffer \gls{pes} when anharmonicity is included, and hence smaller displacements. 
Similar trends, but much less pronounced, are obtained for the nearest-neighbor Ba-Zr and Ba-Ba atomic distances. 

Still, the values including anharmonicity, $\it i.e.$ the SCP and MD values, are larger than the experimental values.
However, using improved approximations for the exchange-correlation approximations within \gls{dft} we expect the agreement to be improved. It was shown in Ref. \cite{granhedBaZrOStabilityPressure2020} that inclusion of Fock exchange as implemented in the two hybrid functionals, HSE \cite{Heyd_Scuseria_Ernzerhof_2003,Heyd_Scuseria_Ernzerhof_2006} and CX0p \cite{Jiao_Schröder_Hyldgaard_2018}, stiffens the atomic bonds and thus increases the vibrational frequencies. It then follows that the \gls{msrd} will decrease.

\begin{table}[h!]
    \centering
    \begin{tabular}{cccccc}\toprule\toprule
                                                                           & T (K)  & Exp    & \gls{md} & \gls{scp} & \gls{hp}\\ \midrule
        \multirow{2}*{$\langle \Delta u^2_{\parallel}\rangle_\text{BaO}$}  &   0   &   -    &   -      & 6.4        &  6.8     \\
                                                                           & 300   & 14.5   &  18.2    & 17.5       &  22.5 \\
        \multirow{2}*{$\langle \Delta u^2_{\parallel}\rangle_\text{BaZr}$} &   0   &   -    &   -      & 2.1        &  2.2   \\
                                                                           & 300   & 6.8    &  7.5     & 7.2        &  8.2  \\
        \multirow{2}*{$\langle \Delta u^2_{\parallel}\rangle_\text{BaBa}$} &   0   &   -    &   -      & 2.7        &  2.7   \\
                                                                           & 300   & 8.7    &  12.3    & 12.0       &  13.5 \\
        \bottomrule\bottomrule
    \end{tabular}
    \caption{Parallel mean squared relative displacements (MSRDs) from \gls{md}, \gls{scp} and \gls{hp},
            in units of 10$^{-3}$ \AA$^2$, based on PBE.
            Compared with experimental data taken from \citet{lebedevStructuralInstabilityBaZrO2013}.}
    \label{tab:MSRD}
\end{table}


\section{Discussion}
\label{sec:discussion}

\subsection{Exchange-correlation functional}

As a compromise between accuracy and computational cost we have based our computations on the PBE functional (see Sec. \ref{subsec:DFT}).
The vibrational properties at the R point are quite sensitive to the exchange-correlation approximation used in the DFT calculations \cite{perrichonUnravelingGroundStateStructure2020}.
To test this sensitivity, we have also computed  the temperature dependent vibrational spectra using the PBEsol functional.

A 4th order \gls{fcp} is constructed based on the PBEsol functional (for details, see Supplemental Material \cite{SI:this}).
In \autoref{fig:rmod_PES_PBEsol} we show the \gls{fcp} result for the PES along the R-tilt mode \gls{fcp} model together with the \gls{dft} data.
In contrast to PBE, the PBEsol functional shows a weak double-well landscape for the R-tilt mode potential.
This means that the R-tilt mode is unstable in the harmonic approximation with an imaginary frequency, in this case, equal to 4.2i\ \SI{}{\milli\electronvolt}. This implies that, in a classical description and using PBEsol, the cubic phase for BZO is unstable close to zero Kelvin.

\begin{figure}[ht]
\centering
\includegraphics{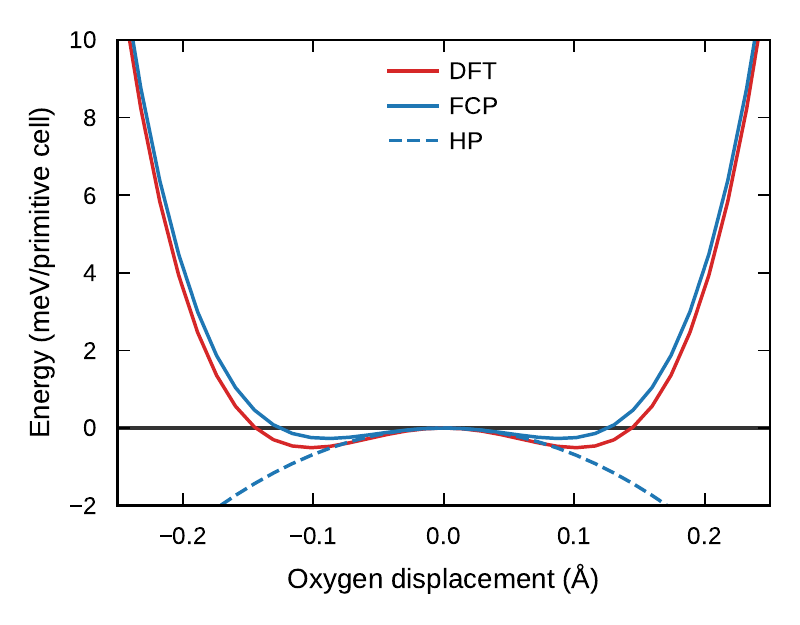}
\caption{
    The potential energy surface (PES) along the R-tilt mode based on PBEsol. 
    The double well minimum potential indicates a structural phase-transition in the classical limit.
}
\label{fig:rmod_PES_PBEsol}
\end{figure}

\begin{figure}[ht]
\centering
\includegraphics{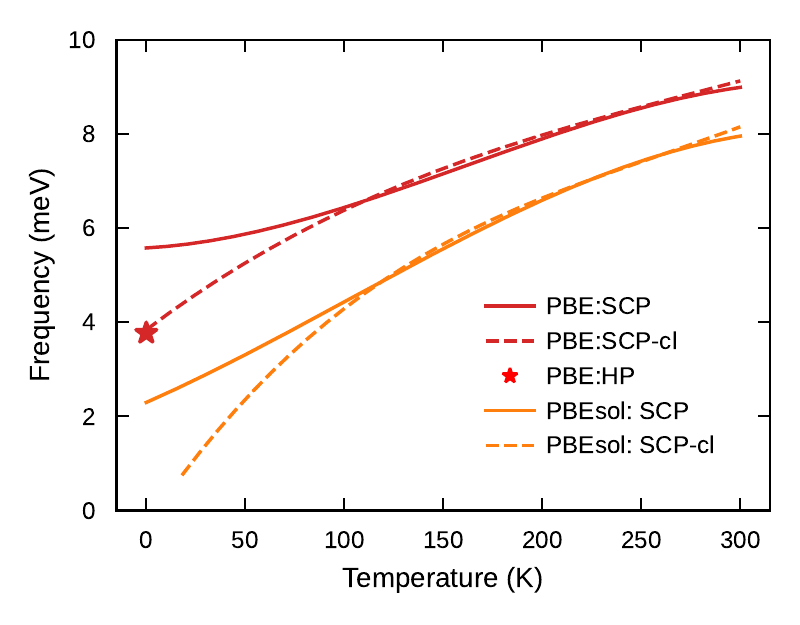}
\caption{
    Theoretical temperature dependence of the R-tilt mode based on PBE and PBEsol, respectively. The \gls{scp}-cl model shows the effect of neglecting the quantum fluctuations of the atomic motion in the \gls{scp} method. The HP is the harmonic approximation.
}
\label{fig:rmod_frequency_PBE_PBEsol}
\end{figure}

The \gls{fcp} based on PBEsol is used in the same way as described for PBE to obtain the temperature dependent phonon frequencies.
In \autoref{fig:rmod_frequency_PBE_PBEsol} we compare our theoretical results using the two different functionals, PBE and PBEsol.
The frequencies based on the PBEsol functional is about \SI{2}{\milli\electronvolt} lower compared with the PBE frequencies and PBEsol gives a somewhat stronger temperature dependence.
Both functionals give rise to a finite frequency at T=\SI{0}{\kelvin}. 
For PBEsol, which has a double-well landscape for the R-tilt mode potential (see \autoref{fig:rmod_PES_PBEsol}), it is worth noting that quantum fluctuations at T=\SI{0}{\kelvin} stabilizes the cubic structure using the SCP model.


\subsection{Effect of quantum fluctuations}

To demonstrate the effect of the quantum fluctuations of the atomic motion on the R-tilt mode we can now compare the result for the \gls{scp} model using the amplitude in \autoref{eq:amplitude_qm}, which contains the proper quantum fluctuations, with the results obtained by using the classical limit of the amplitude,
{\it i.e.},
\begin{equation}
 A_{\lambda}^{i,\textrm{cl}} = \sqrt{{{k_\text B}T}/{(m_i \omega_{\lambda}^2})}\ .
\end{equation}
The latter model is here denoted \gls{scp}-cl and neglects the quantum fluctuations. When lowering the temperature \gls{scp}-cl starts to deviate from \gls{scp} around \SI{100}{\kelvin}, as seen in \autoref{fig:rmod_frequency_PBE_PBEsol}. The frequencies decrease more rapidly compared with the \gls{scp} model and for PBE the frequency approaches the PBE harmonic frequency value of \SI{3.8}{\milli\electronvolt}, the \gls{hp} value. However, the \gls{scp}-cl value for PBEsol approaches zero at a finite temperature. Hence, by neglecting quantum fluctuations, PBEsol predicts an unstable cubic structure at \SI{0}{\kelvin}.

\subsection{Thermal expansion}

We have neglected thermal expansion in our theoretical analysis. In Ref. \cite{perrichonUnravelingGroundStateStructure2020} thermal expansion within the quasi-harmonic approximation was taken into account for the frequency of the R-tilt mode. The PBE value for the frequency increased by \SI{0.8}{\milli\electronvolt} in the temperature interval 0 to 300 K, while the increase for the two hybrid functional HSE and CX0p was less, about \SI{0.5}{\milli\electronvolt}, in the same interval. 
Therefore, the temperature dependence derived here, may increase slightly if thermal expansion is taken into account.

\subsection{Comparison with other work}

Recently Zheng {\it et al.} \cite{Zheng2022} considered anharmonic lattice
dynamics in BZO with focus on thermal transport in the temperature range 300-\SI{2000}{K} using the PBE functional.
They used a non-perturbative self-consistent phonon method as implemented in \textsc{ALAMODE} \cite{TadGohTsu14,tadanoSelfconsistentPhononCalculations2015}.
They also investigated the R-tilt mode and they obtained a frequency increase from \SI{6.3}{\milli\electronvolt} to \SI{8.2}{\milli\electronvolt} in the temperature range 0 to \SI{300}{K} (see their FIG. S4.).
This can be compared with our present SCP results using a similar theoretical approach.
We find an increase from \SI{5.6}{\milli\electronvolt} to \SI{9.0}{\milli\electronvolt} in the same temperature range.
This difference can likely be attributed to the difference in the \gls{fcp}s used in the two different works  (cf. the discussion in Ref. \cite{franssonProbingLimitsPhonon2022}).

Here a 4x4x4 supercell was used to train the \gls{fcp}, whereas Zheng {\it et al.} used a 3x3x3 supercell incommensurate with the R-tilt mode.
Furthermore,  Zheng {\it et al.} constructed the harmonic force-constants separately from the anharmonic force-constants.
These differences in the \gls{fcp} construction likely explains the observed difference in calculated frequencies.

\section{Conclusions}
\label{sec:conclusions}

We have presented a combined experimental and theoretical study of the temperature dependence of the antiferrodistortive R-tilt mode in barium zirconate (BZO). 

Our inelastic neutron and x-ray scattering measurements on a single crystal clearly show that the antiferrodistortive phonon mode at the R point softens substantially, from \SI{9.4}{\meV} at room temperature to \SI{5.6}{\meV} at \SI{2}{\K}. 
In contrast, the barium associated acoustic mode at the same R point is found to be nearly temperature independent.

We use a first-principles computational approach to study the lattice dynamics.
A force constant potential (FCP) including cubic and quartic anharmonic terms is derived using training structures from density functional theory (DFT) calculations.
The PBE functional is employed for the exchange-correlation functional and to gauge the sensitivity with respect to the exchange-correlation functional, the PBEsol functional is also used in some calculations.
The vibrational motion is investigated by direct molecular dynamics (MD) simulations as well as by using a self-consistent phonon approach.
The effect of the anharmonicity on the lattice dynamics is hence taken into account non-perturbatively.

Our theoretical approach, based on a self-consistent phonon approach, predicts a soft phonon mode at the R point, associated with antiphase tilt oscillations of sequential oxygen ZrO$_6$ octahedra, with a strong temperature dependence, from \SI{9.0}{\milli\electronvolt} at \SI{300}{\kelvin} to \SI{5.6}{\milli\electronvolt} at \SI{0}{\kelvin}, in great agreement with the present measurements.
The quantum fluctuations of the atomic motion are found to be important to obtain the proper temperature dependence at low temperature.
To accurately describe the vibrational motion at low temperatures it is thus important to include both anharmonicity and quantum fluctuations of the atomic motion in a consistent way.

The mean squared displacements of the different atoms are determined as function of temperature and are shown to be consistent with available experimental data.
The mean squared relative displacements have also been computed. 
These displacements are measured in EXAFS and the fluctuations of the Ba-O distance are a sensitive local measure of the antiphase tilt oscillations of the oxygen octahedra. 
We compare our computed values for the fluctuation of the Ba-O distance with available EXAFS data at 300 K and the agreement with the EXAFS data is improved when anharmonicity is taken into account in our theoretical modeling.

Altogether, our work provides a robust description of the lattice dynamics and its temperature dependence at the R point, but future work will be needed to clarify open questions on the lattice dynamics away from R. This includes the exact position and temperature dependence of the tilt branch along the R-M direction, as well as a proper deconvolution of the inelastic spectrum. The temperature evolution of the lineshape in INS spectra still needs to be understood and the importance of different ingredients evaluated (resolution effects, anharmonicity, potential presence of nano-domains, etc.). We anticipate that further inelastic scattering experiments or diffuse scattering will help to address this issue and that extended computational modeling may uncover the presence or absence of the proposed nano-domains.
\\

\section{Acknowledgements}
Rapha\"el Haumont and Romuald Saint-Martin from the ICMMO - UMR 8182, Université Paris Saclay, are acknowledged for their help in the preparation of the single crystals used for the INS experiments.
Fredrik Eriksson and Paul Erhart are thanked for useful discussions in connection to the software \hiphive.
Funding from the Swedish Energy Agency (grant No. 45410-1), the Swedish Research Council (2018-06482 and 2020-04935) and the Excellence Initiative Nano at Chalmers is gratefully acknowledged. 
The computations were performed by resources provided by the Swedish National Infrastructure for Computing (SNIC), partially funded by the Swedish Research Council through grant agreement no. 2018-05973.
Cosme Milesi-Brault was supported by the Operational Programme Research, Development, and Education (financed by European Structural and Investment Funds and by the Czech Ministry of Education, Youth, and Sports), Project No. SOLID21-CZ.02.1.01/0.0/0.0/16\_019/0000760.

\end{document}


\title{Supplemental Material: \\ Anharmonicity of the antiferrodistortive soft mode in barium zirconate BaZrO$_3$}

\author{Petter Rosander}
\author{Erik Fransson}
\affiliation{Department of Physics, Chalmers University of Technology, SE-412 96  G\"oteborg, Sweden}

\author{Cosme Milesi-Brault}
\affiliation{Department of physics and materials science, University of Luxembourg, 41 Rue du Brill, L-4422 Belvaux, Luxembourg}
\affiliation{Materials Research and Technology Department, Luxembourg Institute of Science and Technology, 41 rue du Brill, L-4422 Belvaux, Luxembourg}
\affiliation{Institute of Physics of the Czech Academy of Sciences, Na Slovance 1999/2, 182 21 Prague, Czech Republic}
\author{Constance Toulouse}
\affiliation{Department of physics and materials science, University of Luxembourg, 41 Rue du Brill, L-4422 Belvaux, Luxembourg}



\author{Fr\'ed\'eric Bourdarot}
\author{Andrea Piovano}
\affiliation{Institut Laue-Langevin (ILL), 6 Rue Jules Horowitz, 38043 Grenoble, France}
\author{Alexei Bossak}
\affiliation{European Synchrotron Radiation Facility, BP 220, 38043 Grenoble, France}

\author{Mael Guennou}
\email{mael.guennou@uni.lu}
\affiliation{Department of physics and materials science, University of Luxembourg, 41 Rue du Brill, L-4422 Belvaux, Luxembourg}

\author{G\"oran Wahnstr\"om}
\email{goran.wahnstrom@chalmers.se}
\affiliation{Department of Physics, Chalmers University of Technology, SE-412 96  G\"oteborg, Sweden}

\date{\today}

\maketitle

\vspace{1.0cm}
\section{Model validation}
\begin{figure}[ht]
    \includegraphics{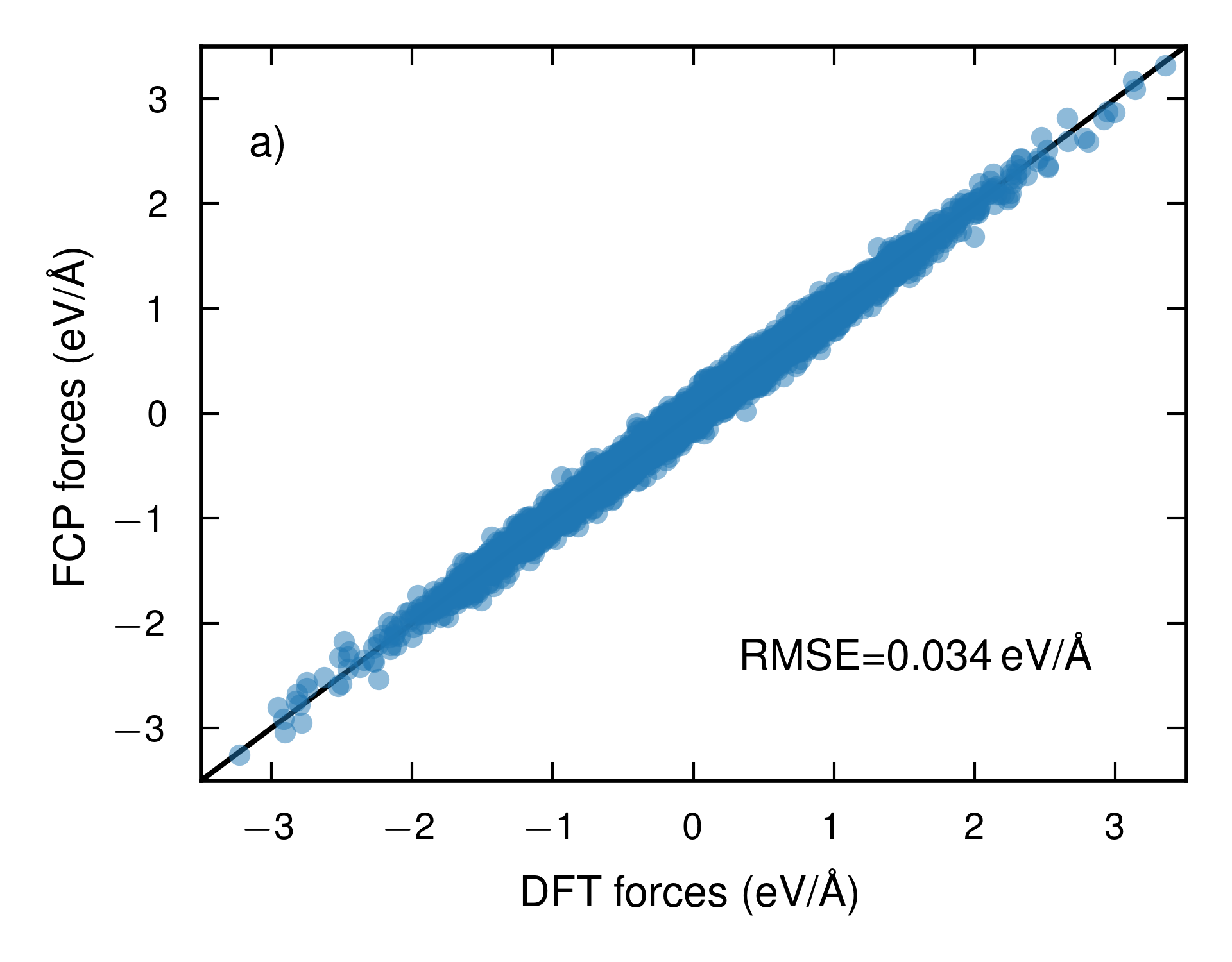}
    \includegraphics{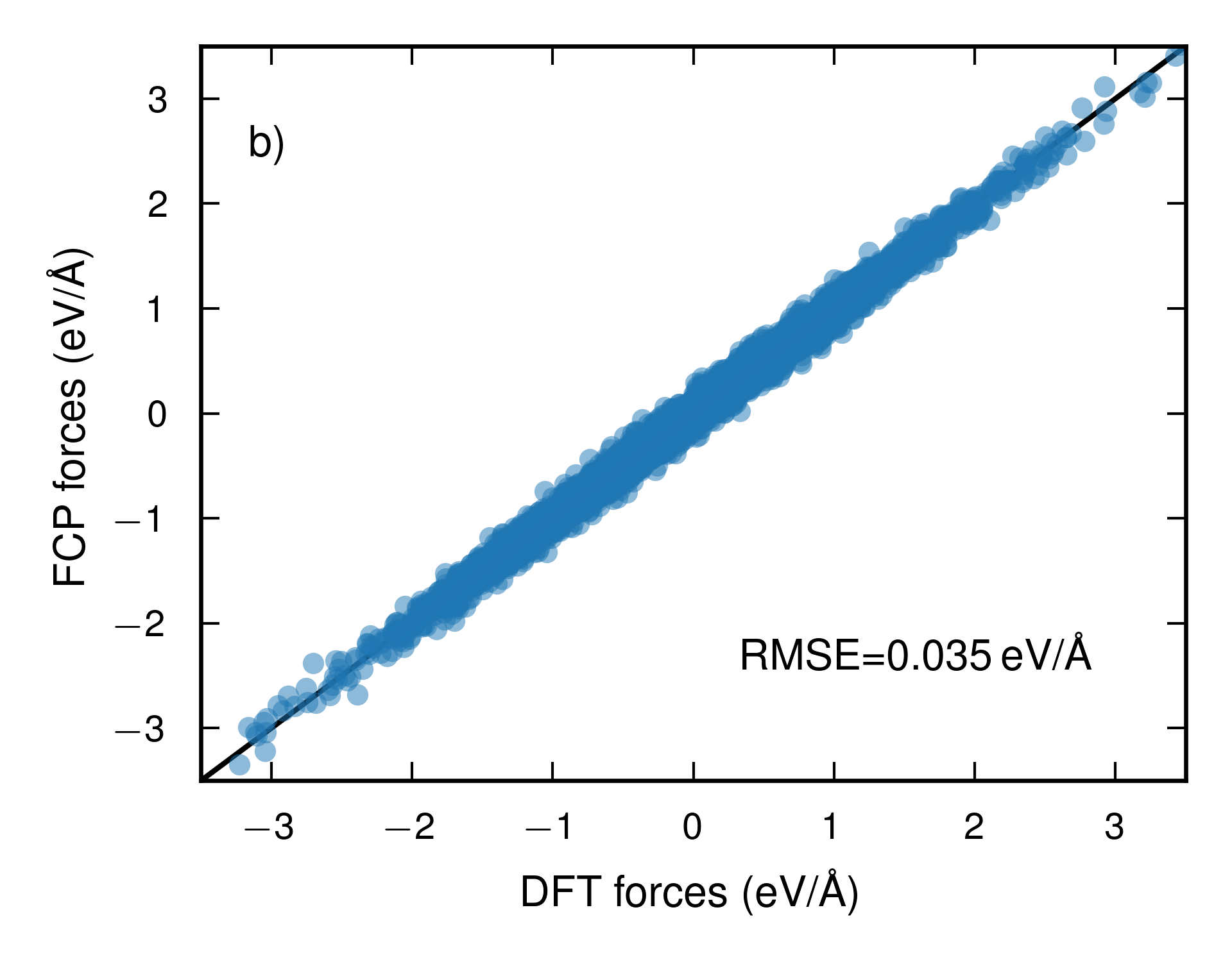}
    \caption{Forces from \gls{dft} compared to predicted forces with the \gls{fcp} for the validation sets from \gls{cv} analysis for a) PBE and b) PBEsol.}
     \label{fig:model_validation_forces}
\end{figure}

The performance of the \gls{fcp} is estimated from a 10-Fold cross validation split.
The parity plot for this analysis is shown in \autoref{fig:model_validation_forces}.
It depicts the \gls{fcp} vs \gls{dft} forces for the validation data.
The obtained average \gls{rmse} over the splits is $\SI{0.034}{eV/Å}$ and $\SI{0.035}{eV/Å}$ for PBE and PBEsol respectively.
This shows that there is a good agreement between \gls{dft} and our \gls{fcp}.

\clearpage
\section{Phonon dispersions}
\begin{figure}[ht]
\centering
\includegraphics[width=.57\textwidth]{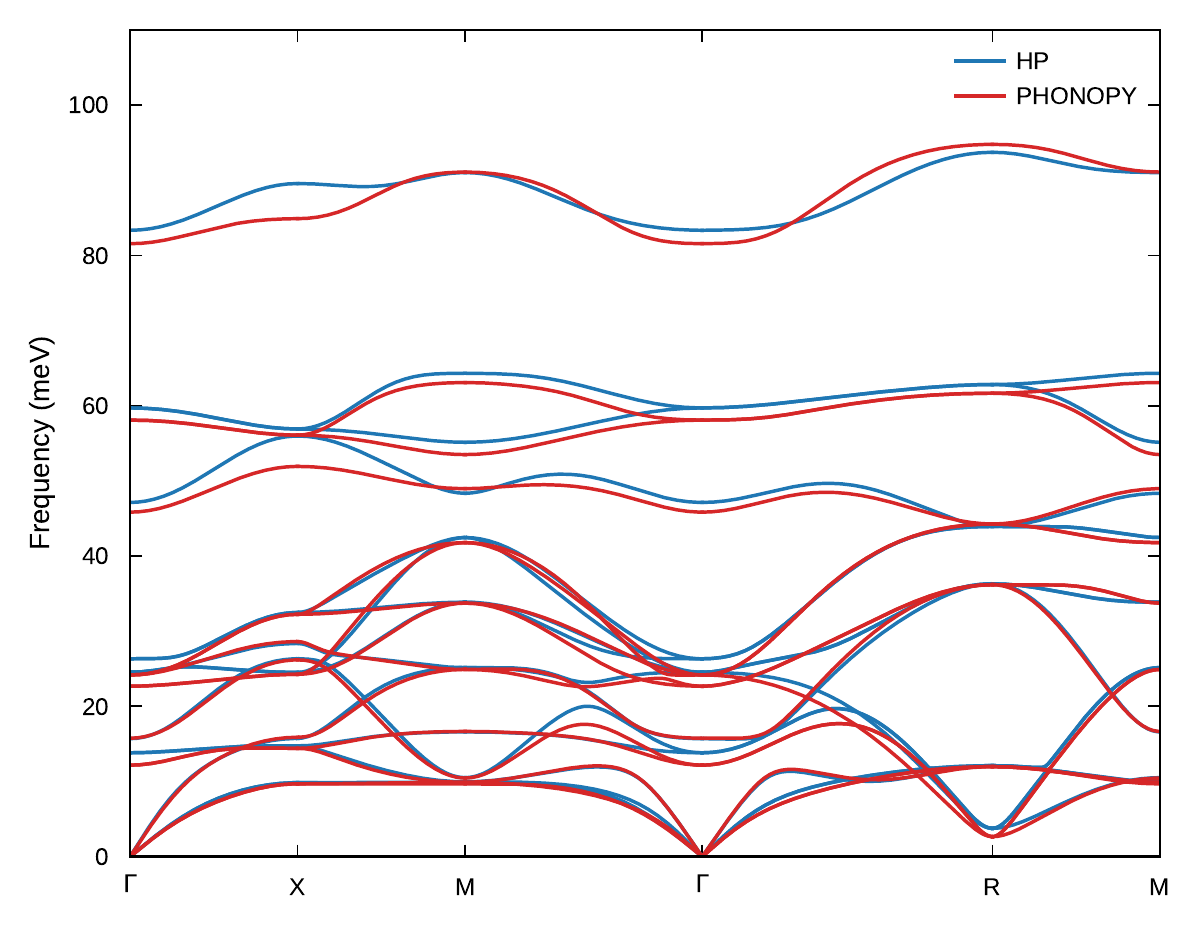}
\caption{
    Phonon dispersions based on PBE. Comparison between our harmonic phonon (\gls{hp}) model based on the derived force constant potential (\gls{fcp}) and the standard small displacement method with $\pm$ 0.01 \AA\ as implemented in \textsc{phonopy}. The R-tilt mode frequency is 3.77 meV and 2.66 meV, respectively. We expect that the deviations at higher frequencies, as seen in the figure, can be reduced by using more \gls{dft} training configurations and possibly a higher order expansion.
}
\label{fig:compare-dispersion-with-phonopy}
\end{figure}

\begin{figure}[ht]
\centering
\includegraphics[width=.57\textwidth]{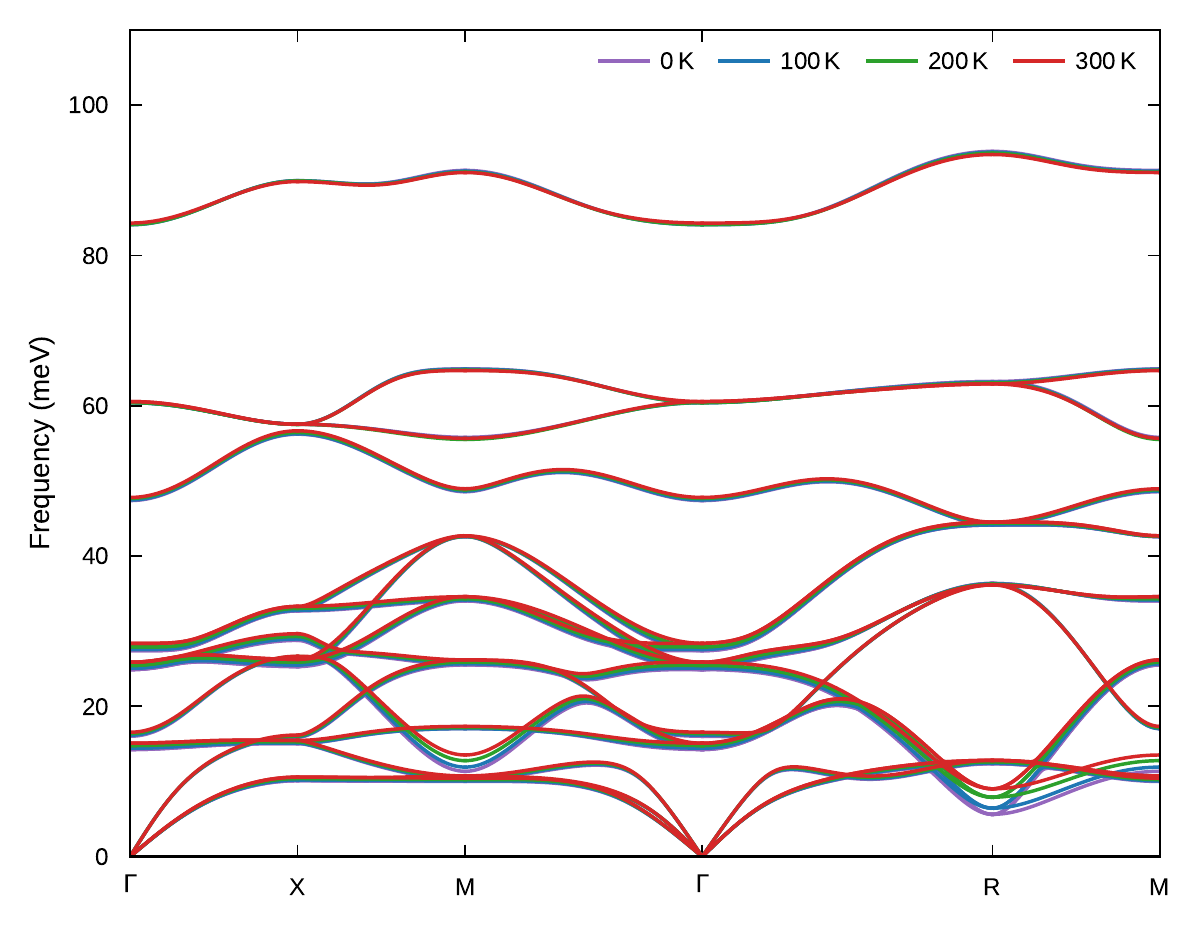}
\caption{
    Phonon dispersions based on PBE. The data are obtained using the self consistent phonon (\gls{scp}) method using the derived force constant potential (\gls{fcp}). Data are shown at four different temperatures; 0, 100, 200 and 300 K. It is seen the temperature dependence is mainly localized at the R point (out of phase tilting) and the M point (in phase tilting). At the R point the tilting frequency is increased from 5.63 meV to 9.03 meV when increasing the temperature to 300 K.
}
\label{fig:compare-dispersion-with-phonopy}
\end{figure}

\section{Details for fitting of inelastic neutron scattering spectra}

\begin{figure}[ht]
    \centering
    \includegraphics[width=.65\textwidth]{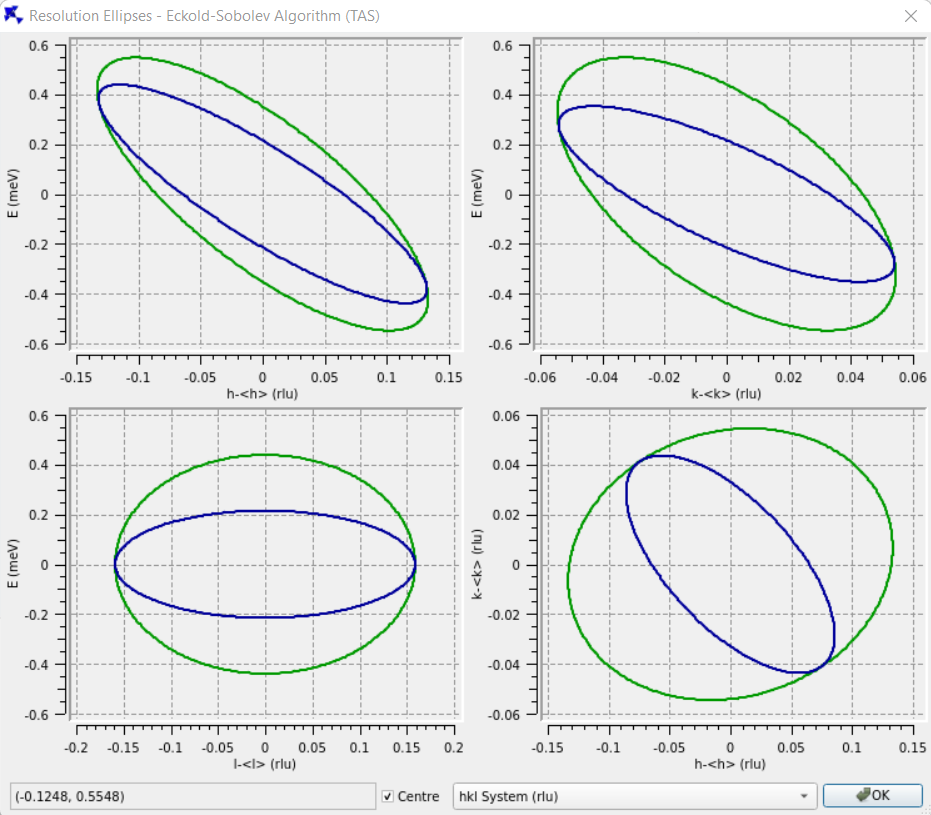}
    
    \caption{Ellipsoids of resolution based on the TAS file from the Institut Laue Langevin (ILL) IN8 spectrometer. The green line represents HWHM contour of the projected ellipse while the blue one represents the HWHM contour of the sliced ellipse.}
    \label{fig:ellipsoid}
\end{figure}

Fitting of the inelastic neutron scattering (INS) data is performed by using the Takin software\cite{Weber2016, Weber2017, Weber2021} using damped harmonic oscillators. This takes into account the resolution function of the diffraction setup (cf.\ the resolution ellipsoids on Fig. \ref{fig:ellipsoid}). In a first step, a parabolic dispersion around the R point was introduced for the tilt mode in a attempt to fit properly the asymmetry of this peak at low temperatures. This unfortunately was not successful, probably because of the very steep dispersion. We therefore added an additional ad-hoc mathematical asymmetry $\gamma$ to the damping $\Gamma$, i.e. $\Gamma=2\Gamma_0 /(1+e^{\gamma (E-E_0)})$, for the damped harmonic oscillator of lowest energy. This allowed us to obtain reasonable fits for the mode energies.


\clearpage
\section{Dispersion data}
\subsection{Dispersion from INS spectra at 2 K}

\begin{figure}[h]
    \centering
    \includegraphics[height=.65\textwidth]{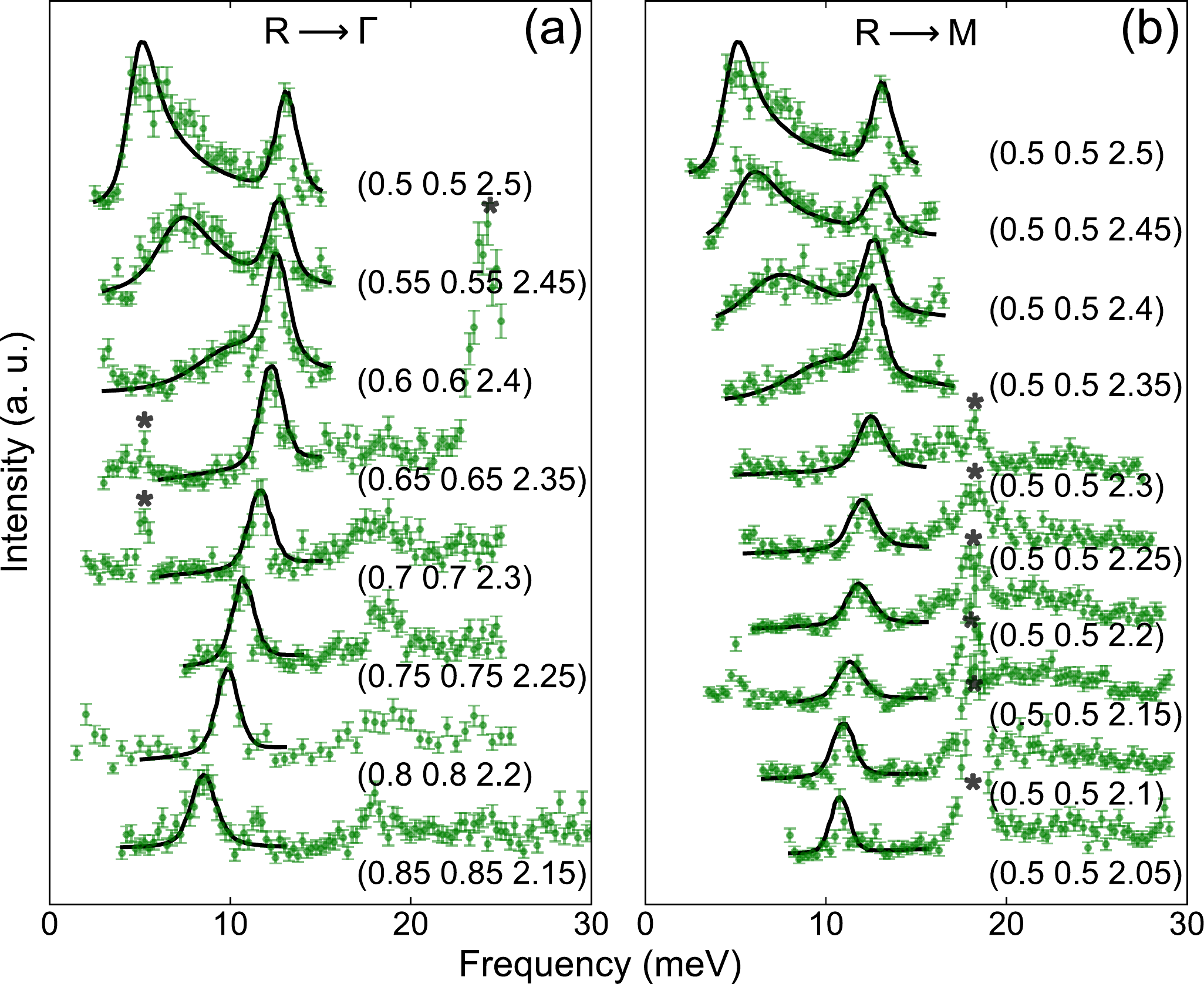}
    \caption{INS spectra taken at 2 K along the (a) R-$\Gamma$ and (b) R-M directions of the Brillouin zone. Black lines show fits used to plot the dispersion of Fig.~5 of the main text. Stars ($\ast$) indicate spurious peaks. Their spurious nature has every time been checked by measuring a spectrum at an equivalent point in a different Brillouin zone.} 
    \label{fig:INS-spectra}
\end{figure}



\clearpage
\subsection{Dispersion from IXS spectra at 80 K}

\begin{figure}[htpb]
    \centering
    \includegraphics[height=.65\textwidth]{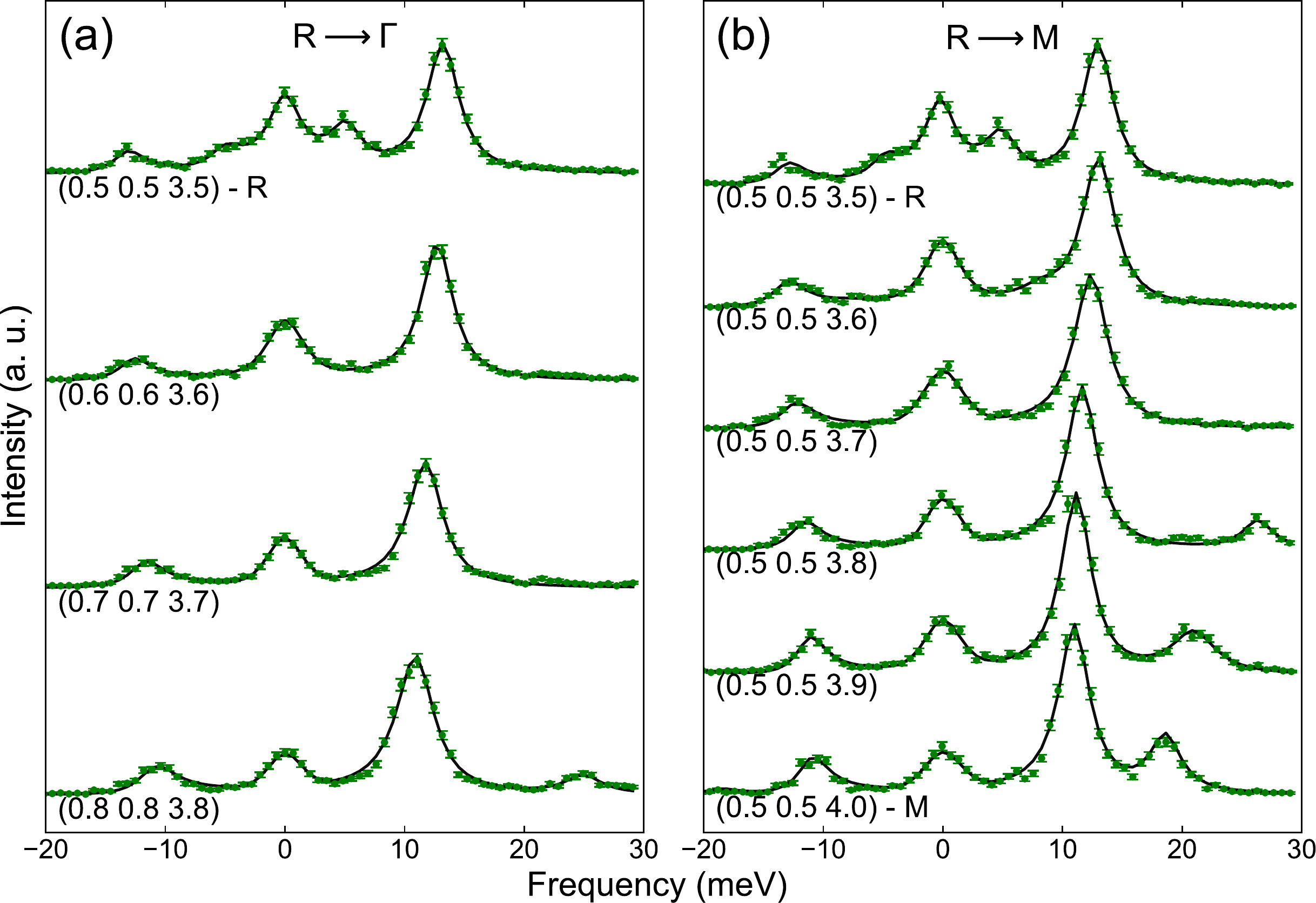}
    \caption{IXS spectra taken at 80 K along the (a) R-$\Gamma$ and (b) R-M directions of the Brillouin zone. Black lines represent fits used to extract frequency of the R-acoustic mode, shown on the dispersion curves of Fig.~5 in the main text.}
    \label{fig:IXS_dispersion}
\end{figure}


\clearpage
\section{Peak fitting}

\begin{figure}[ht]
\centering
\includegraphics[width=.45\textwidth]{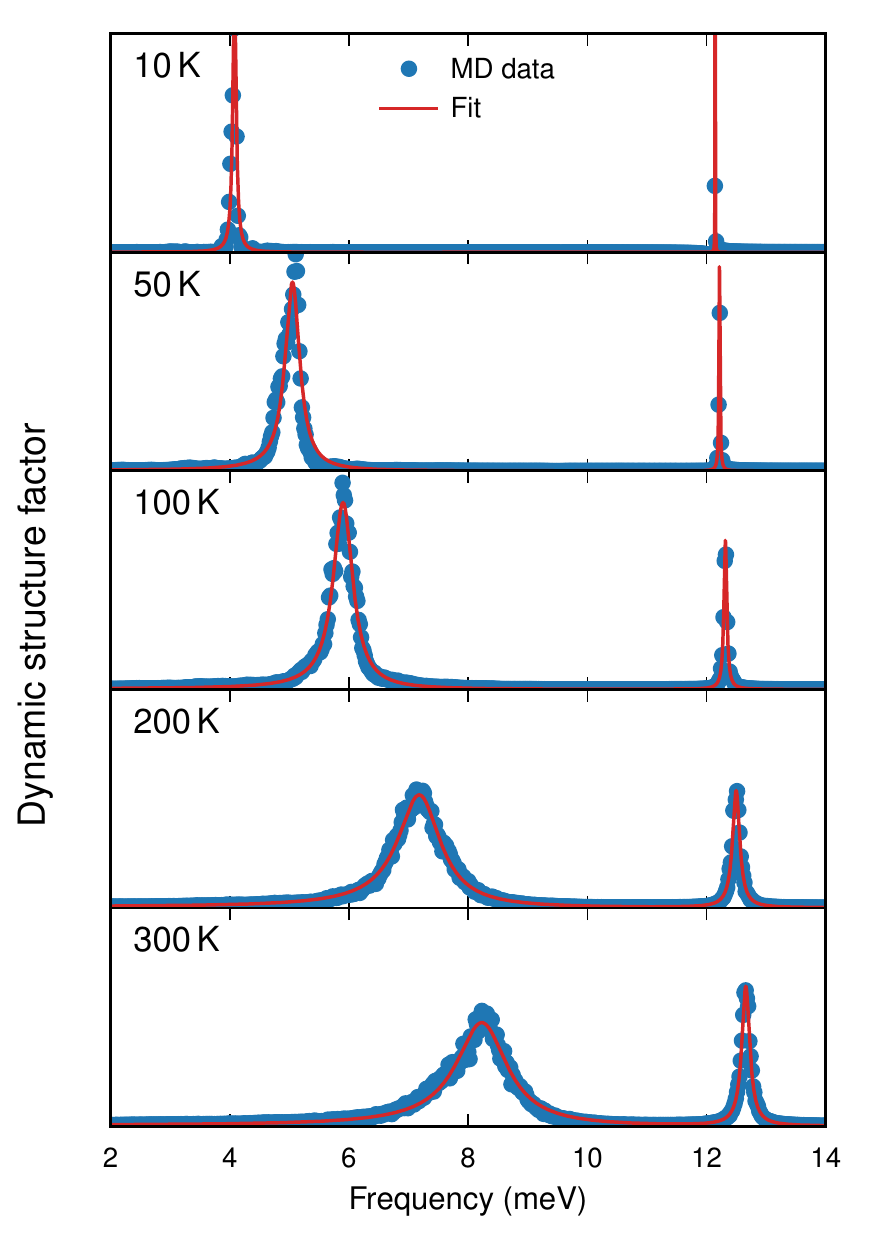}
\caption{
    The dynamical structure factor $S(\boldsymbol{q},\omega)$ at different temperatures for PBE. The damped harmonic oscillator model in \autoref{eq:damped} is fitted to the MD data.
}
\label{fig:peak_fitting}
\end{figure}

The calculated dynamical structure factor, $S(\boldsymbol{q},\omega)$, from the MD simulation is fitted to a damped harmonic oscillator model
\begin{equation}
    f(\omega) = A\ \frac{2\Gamma\omega_0^2}{(\omega^2 - \omega_0^2)^2 + (\Gamma\omega)^2}\ ,
    \label{eq:damped}
\end{equation}
where $\omega_0$ is the bare frequency, $\Gamma$ the damping and $A$ the amplitude of the mode \cite{FraSlaErhWah2021}.
These fits are shown in \autoref{fig:peak_fitting} for different temperatures. In \autoref{tab:SQ} we show the fitted values for the bare frequency $\omega_0$ and the damping coefficient $\Gamma$. The frequency at the maximum of the peak is given by $\omega_{\text{max}} = \omega_0 \sqrt{1 - \Gamma^2/(2 \omega_0^2})$.

\begin{table}[h!]
    \centering
    \begin{tabular}{ccccc}\toprule\toprule
                                        & \multicolumn{2}{c}{R-tilt} & \multicolumn{2}{c}{R-acoustic} \\
                                  T (K) & $\omega_0$ (meV) & $\Gamma$ (meV)      & $\omega_0$ (meV) & $\Gamma$ (meV)          \\ \midrule
                                  10    &   $4.08$   &  $0.07$       &  $12.14$   &  $0.00$           \\
                                  50    &   $5.06$   &  $0.30$       &  $12.22$   &  $0.02$           \\
                                 100    &   $5.91$   &  $0.40$       &  $12.31$   &  $0.06$           \\
                                 200    &   $7.21$   &  $0.87$       &  $12.50$   &  $0.14$           \\
                                 300    &   $8.26$   &  $1.04$       &  $12.66$   &  $0.16$          \\
        \bottomrule\bottomrule
    \end{tabular}
    \caption{Fitted values for the bare frequency $\omega_0$ and the damping coefficient $\Gamma$ from the damped harmonic oscillator in \autoref{eq:damped} for the dynamical structure factor from the MD simulation.}
    \label{tab:SQ}
\end{table}


\clearpage
\section{Frequencies obtained from an effective harmonic model}
\begin{figure}[ht]
\centering
\includegraphics[width=0.6\textwidth]{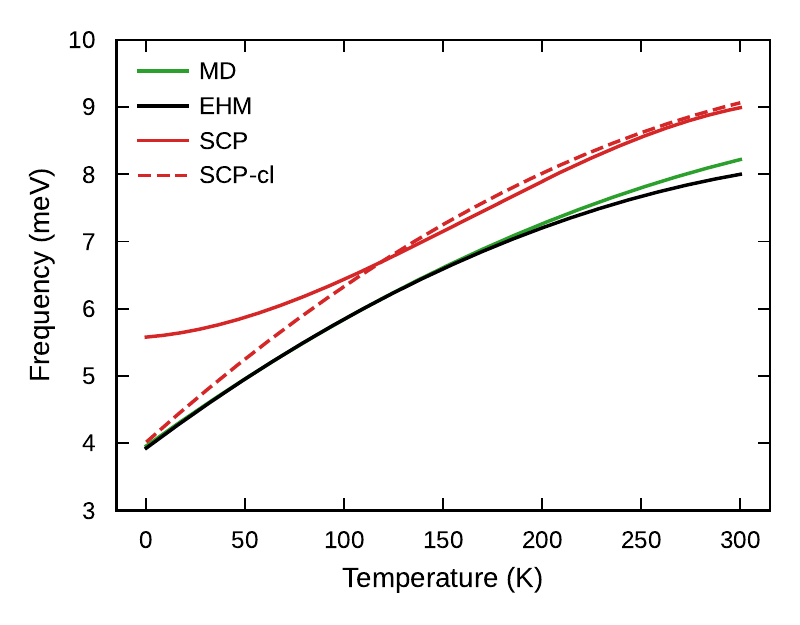}
\caption{
    Calculated frequencies for four different models using PBE.
    Only the SCP model includes the quantum fluctuations of the atomic motions.
}
\label{fig:MD-EHM-freqs}
\end{figure}

It is possible to construct an \gls{ehm} directly from the \gls{md} simulation \cite{hellmanLatticeDynamicsAnharmonic2011}.
This can be done by minimizing the differences between the forces for the harmonic model and the \gls{fcp} in the MD simulation,
\[
    \min_{\mathbf{x}}\lVert {\mathbf{A}}(\mathbf{u}) {\mathbf x} - {\mathbf{f}}(\mathbf{u}) \rVert,
\]
which is the same minimization problem as for the \gls{scp} problem.
Hence, the only difference between the two methods is how the displaced structures are obtained.
The frequency obtain from the EHM method is compared with the frequencies presented in the paper.
It compares very well with the bare frequency of the antiferrodistortive R-tilt mode from the dynamical structure factor (the MD model) as can be seen in \autoref{fig:MD-EHM-freqs}.
The fact that the \gls{scp} gives a slightly larger frequency compared to the \gls{ehm} has been observed in previous studies as well, e.g., in Ref.~
\cite{korotaevReproducibilityVibrationalFree2018,metsanurkSamplingdependentSystematicErrors2019,castellanoInitioCanonicalSampling2022,franssonProbingLimitsPhonon2022}
